\documentclass[vecphys]{svmult}

\usepackage{makeidx}         
\usepackage{graphicx}        
\usepackage{multicol}        
\usepackage{cite}            
\usepackage[bottom]{footmisc}

\makeindex             

\begin{document}

\title{Spin Ice}
\author{Michel J.P. Gingras}
\institute{Department of Physics and Astronomy,
University of Waterloo, \\ Waterloo, Ontario, N2L-3G1, Canada \\
and\\
Canadian Institute for Advanced Research/Quantum Materials Program \\
180 Dundas Street West, Suite 1400, Toronto, Ontario, M5G-1Z8, Canada\\
\texttt{gingras@gandalf.uwaterloo.ca}
}

\maketitle


\abstract

Geometric frustration usually arises in systems that comprise 
magnetic moments (spins) which reside on the sites of a lattice 
made up of elementary triangular or tetrahedral units and which 
interact via antiferromagnetic nearest-neighbor exchange.
Albeit much less common, geometric frustration can also arise
in systems with strong non-collinear single-ion easy-axis (Ising-like)
anisotropy and {\it ferromagnetically} coupled spins.
This is what happens in some pyrochlore oxide materials where
Ising-like magnetic rare earth moments (Ho$^{3+}$, Dy$^{3+}$)
sit on a lattice of corner-shared tetrahedra and are coupled 
via effectively ferromagnetic (dipolar) interactions.
These systems possess a macroscopic number of quasi-degenerate 
classical
ground states and display an extensive low-temperature entropy
closely related to the extensive proton disorder entropy in 
common water ice. For this reason, these magnetic systems are
called {\it spin ice}. This chapter reviews the essential 
ingredients of spin ice phenomenology in magnetic pyrochlore
oxides.

\section{Introduction}

In some geometrically frustrated magnetic systems, 
there exists an exponentially large number, $\Omega_0$, 
of degenerate classical ground states.
This gives rise to an extensive residual, or zero point, ground state entropy, 
$S_0=k_{\rm B}\ln(\Omega_0)$. 
The most celebrated example of such a system 
is the triangular lattice with Ising spins interacting via nearest-neighbor 
antiferromagnetic exchange. Indeed, as shown by Wannier in 1950,
this system remains disordered at all temperatures, with  a ground state that
has two spins up and one spin down (or vice versa) per triangle and has a zero
point entropy per site $S_0\approx 0.323$ $k_{\rm B}$ \cite{Wannier:1950}.
However, the triangular Ising antiferromagnet was not the first
condensed matter system identified as having a residual entropy. 
In fact, the first such system was not even 
a magnetic one. Fifteen years or so earlier, William Giauque
(Chemistry Nobel Prize, 1949)
and co-workers had performed thermodynamic measurements and determined
that the solid phase of common water ice possesses an unaccounted 
residual entropy~\cite{Giauque:1933,Giauque:1936}.
This result was soon explained by Linus Pauling 
(Chemistry Nobel Prize, 1954) in terms of a macroscopic number of
proton (H$^+$) configurations in water ice arising from
the mismatch between the crystalline symmetry of ice and the 
local hydrogen bonding requirement of the water molecule~\cite{Pauling:1935}.

Over the past ten years, a certain class of insulating magnetic 
materials in which the configurational disorder in the 
orientations of the magnetic moments is precisely the same as that of water ice have
been the subject of numerous experimental and theoretical studies.
Because of their analogy with water ice, 
these systems have been coined the name {\it spin 
ice}~\cite{Harris:1997,Harris:JMMM1998,Ramirez:Nature1999,Bramwell:Science,BGH}.
Most chapters in this book focus on geometrically frustrated antiferromagnets.
The main reason for the interest in frustrated antiferromagnets 
is the pursuit of novel quantum ground states with exciting properties 
which, because of the increased quantum zero point motion caused by the frustration, 
lack conventional semi-classical long-range N\'eel order. 
This chapter differs in that the spin ices are 
frustrated Ising {\it ferromagnets} 
and where quantum fluctuations do not play a significant role. 
Yet, experimental and theoretical studies have
revealed a great richness of equilibrium and non-equilibrium
thermodynamic behaviors in spin ice systems~\cite{BGH}.
This chapter reviews some of the salient elements of the spin ice phenomenology.
It draws particular attention to the problem of water ice and the
semi-formal origin of the Ising nature of the magnetic moments in
spin ice materials, two topics that are not usually covered in detail in
standard graduate solid state textbooks.
It also reviews in some detail the
mean-field theory of spin ices as this
simple tool played a key role in uncovering the microscopic 
origin behind the emergence of the {\it ice rules} in real
dipolar spin ice materals.
We end the chapter with a brief discussion of 
research topics on spin ices that are of current interest.

\begin{figure}[ht]
\begin{center}
\includegraphics[height=3cm,width=3cm,angle=0]{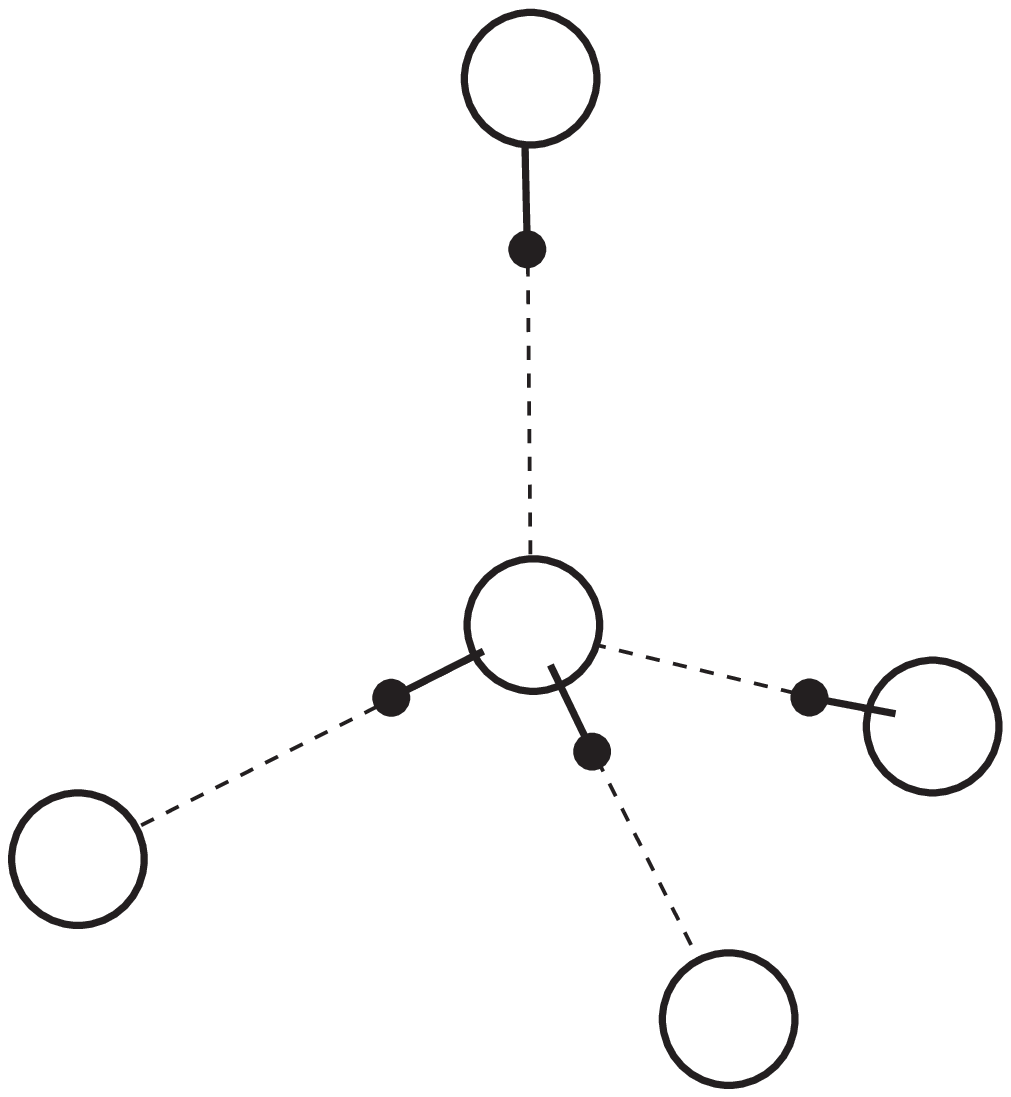}
\includegraphics[height=3cm,width=1.5cm,angle=0]{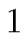}
\includegraphics[height=3cm,width=3cm,angle=0]{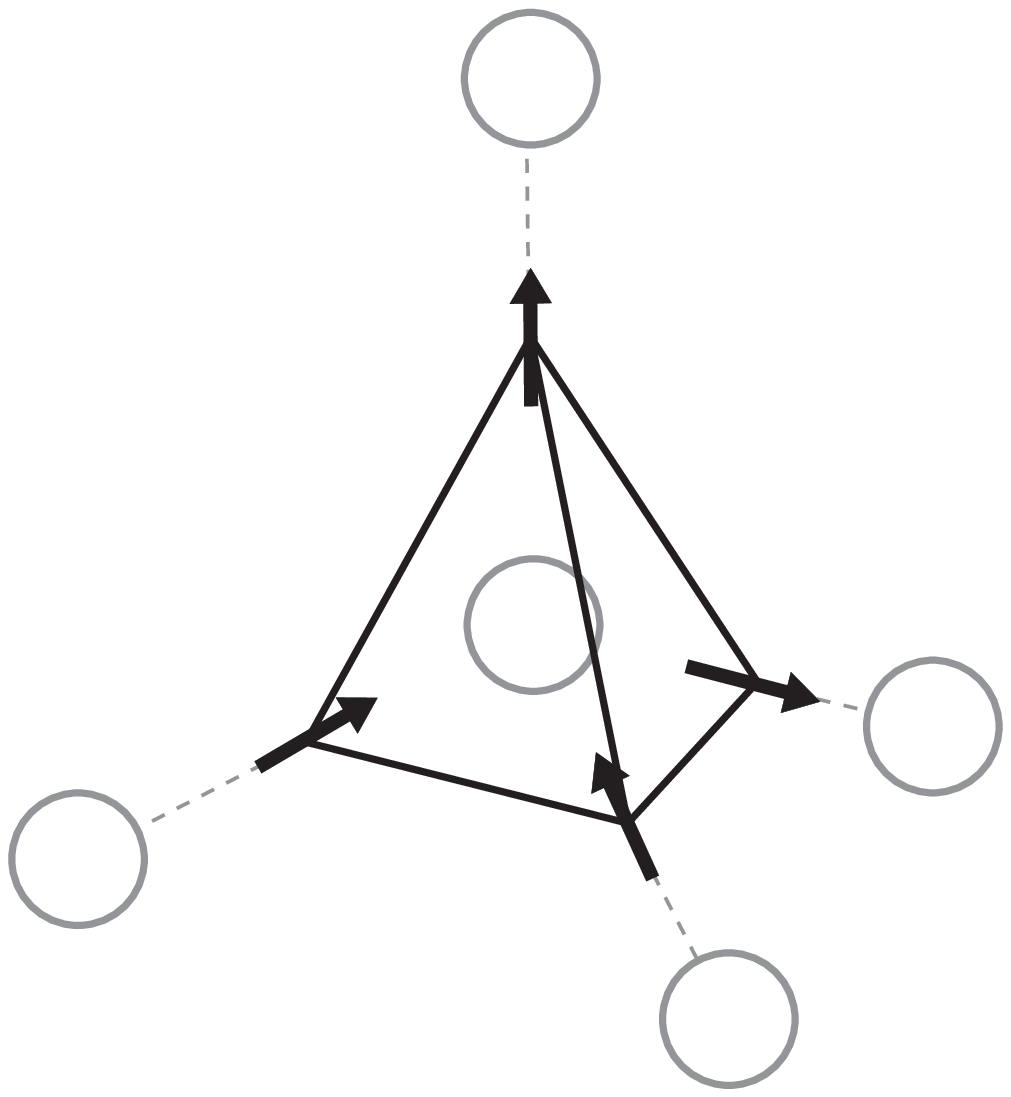}
\caption{Left:
Local proton arrangement in water ice, showing O$^{2-}$ ions
(large white circles) and protons (hydrogen ions, H$^+$, small black circles).
 Each O$^{2-}$ is tetrahedrally coordinated with four other O$^{2-}$, 
with two near covalently bonded protons, 
and two further hydrogen bonded protons.  
In the hexagonal phase of ice, I$_{\rm h}$,
the low energy configurations obey the
so-called Bernal-Fowler ``ice rules''~\cite{Bernal:1933} 
where each O$^{2-}$ oxide
has ``two-near'' and ``two-far'' protons. 
Right: Same as left picture, but where 
the position of a proton is 
represented by a displacement vector (arrow) located at the mid-point
of the O$^{2-}$$-$O$^{2-}$ (oxide-oxide) bond.
The ``two-near/two-far'' Bernal-Fowler ice rule then 
translates into a ``two-in/two-out'' configuration
of the displacement vectors. 
The displacement vectors become the Ising magnetic moments in spin ice
(see Fig.~\ref{pyrochlore}). }
\label{spinice}
\end{center}
\end{figure}

\section{Water Ice, Pauling's Entropy and Anderson's Model}

\subsection{Water ice and Pauling's model}

Water ice is a fascinating strongly correlated condensed matter system,
not least because it  exhibits a seemingly violation
of the third law of thermodynamics. 
In the early 1930s, a series of remarkable specific heat 
experiments by  William Giauque and co-workers found 
that the limiting low-temperature state of
the common (hexagonal) state of water ice, referred to as 
I$_{\rm h}$,  is characterized by a residual entropy,
$S_0 = 0.82 \pm 0.05$ Cal/deg$\cdot$mol, that differs from zero
by a value
far in excees of experimental errors~\cite{Giauque:1933,Giauque:1936}.
In a famous 1935 paper, Linus Pauling showed that, 
because of its configurational proton disorder, I$_{\rm h}$
possesses a finite entropy at zero temperature 
estimated as 0.81 Cal/deg$\cdot$mol~\cite{Pauling:1935}, 
hence very close to the experimental value.

The ice problem is a classic example of 
how the separation of energy scales in an interacting system
can leave some effective low energy degrees of freedom 
innoperative and ultimately frustrated in 
a system's overall agenda in minimizing its energy at low temperature
via local dynamical processes and over a finite, as opposed
to infinite, amount of time.
Here, the chemical binding energy of the water molecule
is so strong, 221 kCal/mol, 
that its chemical structure is left essentially unaltered in the solid phase.
Consequently, the ground state of water ice does not, or more precisely,
dynamically cannot, minimize the electrostatic energy of a globally 
neutral ensemble of ${\rm O^{2-}}$ and ${\rm H^+}$ ions. 
Rather, in the hexagonal (``wurtzite'') and cubic (``sphalerite'') phases of 
ice~\cite{Ziman:1976}, the O$^{2-}$ ions form
an open tetrahedral structure, whose $109$ degree angles 
accomodate almost perfectly that of 
the ${\rm H-O-H}$ bonds of an isolated H$_2$O molecule. 
In the wurtzite phase, the bond length between two distinct O$^{2-}$ ions is 
$2.76$ \AA, while the covalent ${\rm O-H}$ bond of a H$_2$O molecule is a much 
smaller $0.96$ \AA. As the integrity of the 
molecular structure of H$_2$O is maintained
in the solid phase, the minimum energy position of a H$^+$ proton 
is not at the midpoint between two  O$^{2-}$ ions.  
Instead, there are
two equivalent positions for a proton lying on an ${\rm O}-{\rm O}$
bond.
With the four-fold oxygen co-ordination in the hexagonal wurtzite structure,
there is in effect one proton per ${\rm O}-{\rm O}$ bond on average. 
The constraint imposed by the energetically robust
${\rm H_2O}$ structure
therefore results in the two so-called Bernal-Fowler {\it ice rules}
which govern what are acceptable low-energy proton configurations
in the hexagonal wurtzite structure ~\cite{Bernal:1933}.
The first ice rule states that
there should only be one proton per ${\rm O}-{\rm O}$ on average.
The second rule states that, in order for ice to consist
of a hydrogen-bonded solid of water molecules, 
for each O$^{2-}$ ions, two H$^+$ protons
must be in a ``near-to-O$^{2-}$'' position, 
and two protrons must be in a ``far position'' (see Fig.~\ref{spinice}). 
From a solely electrostatic point of view, the protons would like to be as far
apart as possible. The ice rules, implemented via the unaltered integrity
of the H$_2$O water molecule, render the proton-proton interaction an
effectively low-energy part of the problem, and frustrates it.

The ice rules were put forward by Bernal and Fowler in 1933~\cite{Bernal:1933}.
At that time, X-ray diffraction could only determine the 
lattice structure for the oxygen atoms.
According to the first ice rule, 
Bernal and Fowler argued that the hydrogens must lie 
along the direct oxygen-oxygen line of contact (bond) 
between two H$_2$O molecules.
They proposed a regular crystalline proton structure, 
presumably expecting that this would be the case.
However, around the same time, Giauque and co-workers had already obtained 
compelling evidence for a residual zero point entropy in water ice~\cite{Giauque:1933}. 
This led Pauling to his proposal, published in 1935, 
that the open tetrahedral structure of ice leads 
to many equivalent ways of satisfying the Bernal-Fowler ice rules, 
and hence to an extensive entropy~\cite{Pauling:1935}.

Pauling put forward an elegant 
argument to estimate the configurational proton entropy. 
The argument goes as follows. First
consider one mole of ice containing $N_0$ 
${\rm O^{2-}}$ ions and, therefore, $2N_0$ 
${\rm O-O}$ bonds for the hexagonal structure of ice in which 
no two proton are on any given ${\rm O-O}$ bond. That is,
all bonds are taken to obey the first Bernal-Fowler ice rule.
Each ${\rm O-O}$ bond can be taken as having two possible 
positions for a proton. 
This gives $2^{(2N_0)}$ possible proton configurations for the whole system.
Out of the 16 possible configurations associated with each oxygen,
ten are energetically unfavourable:
the ${\rm OH_4^{2+}}$
configuration, the 4 ${\rm OH_3^{+}}$ configurations, the 4 ${\rm OH^{-}}$
configurations and the ${\rm O^{2-}}$ configuration.
This leaves six configurations that satisfy the
Bernal-Fowler rules as the allowed local proton configurations
around each oxygen ion. An upper bound on the number of
ground state configurations, $\Omega_0$,
can therefore be 
estimated by reducing the above $2^{2N_0}$ configurations by a simple 
$6/16$ deflation weight  factor for each oxygen ion.  
This gives $\Omega_0 \leq 2^{2N_0}(6/16)^{N_0}= (3/2)^{N_0}$. 
The corresponding configurational entropy, 
$S_0=k_{\rm B}\ln(\Omega_0) = N_0k_{\rm B}
\ln(3/2)=0.81$ Cal/deg$\cdot$mol,
 is in excellent agreement with the residual entropy 
of  $0.82 \pm 0.05$ Cal/deg$\cdot$mol
determined by Giauque and Stout~\cite{Giauque:1936}.
Pauling's calculation neglects the global constraint on the
number of protons as well as the local constraints coming from closed
loops on the wurtzite lattice. Nevertheless, Pauling's 
estimate has been shown to be accurate to $1-2\%$~\cite{Nagle:1966}.

\subsection{Cation ordering in inverse spinels and
antiferromagnetic pyrochlore Ising model}

\label{cation-ordering}

\begin{figure}[ht]
\begin{center}
\includegraphics[height=6cm,width=7.4cm,angle=0]{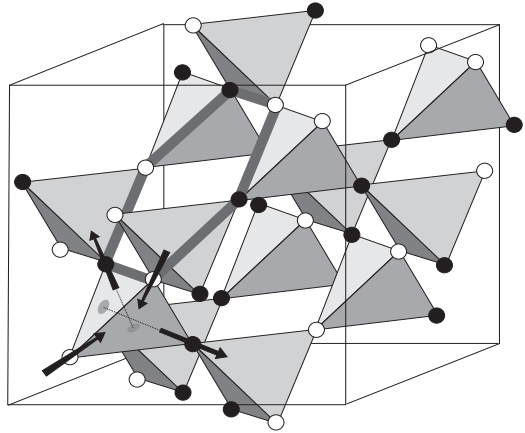}
\caption{
Pyrochlore lattice of corner-sharing
tetrahedra, as occupied by Ho$^{3+}$ and Dy$^{3+}$ in 
in the Ho$_2$Ti$_2$O$_7$ and Dy$_2$Ti$_2$O$_7$ spin ice materials.
The magnetic rare-earth 
Ising moments occupy the corners of the
tetrahedra, as shown on the lower left ``downward'' tetrahedron of the
lattice (arrows). The spins here are the equivalents of the
proton displacement vectors in Fig.~\ref{spinice}.
Each spin axis is along its 
local $[111]$  quantization axis, which goes from one
site to the middle of the opposing triangular face (as shown by the
disks) and meets with the three other $\langle 111 \rangle $  axes in
the middle of the tetrahedron. In the spin ice
materials,
 the ``two-in/two-out''
condition arises from the combined effect of magnetic
(exchange and dipole-dipole) interactions and the strong
Ising anisotropy.
For clarity, other spins on the lattice are denoted by
black and white circles, 
where white represents a spin pointing into a downward
tetrahedron while black is the opposite. 
The entire lattice is shown in an ice-rules state 
(two black and two white sites for every
tetrahedron).
The hexagon (thick gray line) pertains to the loop
excitations used in the ``loop Monte Carlo'' simulations
discussed in Section~\ref{Section:loopMC}.
The shown hexagon is the  smallest
possible loop move involving multiple spins, and which corresponds to
reversing all spins on the loop to produce a new ice-rules obeying 
state.  The reversal of spins on closed loops
 are the lowest-energy excitations that 
allow the system to explore the quasi-degenerate ice rule
manifold of dipolar spin ice at low temperatures.
(Figure reprinted with permission from
R. ~G. Melko {\it et al.},
Phys. Rev. Lett. \textbf{87}, 067203 (2001).
Copyright 2001 by the American Physical Society.)
}
\label{pyrochlore}
\end{center}
\end{figure}

In a 1956 paper~\cite{Anderson:1956}, 
Anderson investigated the problems of cation ordering
in the so-called inverse spinel materials and that of magnetic ordering in 
normal spinels. To a first approximation, both of these problems map
onto the problem of an Ising model with antiferromagnetic nearest-neighbor 
exchange interaction on the B-site of the spinel lattice. This lattice
is structurally identical to the pyrochlore lattice shown in Fig.~\ref{pyrochlore}.
For a discussion of magnetic ordering in spinels, see the Chapter by Takagi in this book.

Consider first the problem of cation ordering in the inverse spinel.
The associated minimum energy problem consists 
in placing $+$ and $-$ signs on the pyrochlore lattice 
(see Fig.~\ref{pyrochlore}) in a ratio 1:1 
such that the number of $+/-$ pairs is maximized. This condition
is satisfied with two $+$ and two $-$ signs for each tetrahedron.
The center of the tetrahedra in the spinel lattice are arranged with 
respect to each other on the same lattice at the oxygen ions in the cubic phase
of water ice (I$_{\rm c}$).
As shown in Fig.~\ref{pyrochlore}, 
the spinel latice and, equivalently, the pyrochlore lattice,  
consists of alternating ``upward'' and ``downward'' tetrahedra. 
Consequently, the problem of $+$ and $-$ charge organization problem
maps onto an ice rule problem if a $+$ sign corresponds to a ``proton near'' on
an upward tetrahedron and a ``proton far'' on a downward tetrahedron, and vice-versa
for a $-$ sign. Neglecting the same constraints that Pauling neglected
for the water ice problem, the problem of cation ordering in inverse spinels
therefore maps onto an ice-like problem, and is therefore characterized by
a Pauling zero point entropy~\cite{Anderson:1956}. 
The problem of cation ordering is also manifestly
the same as that of an antiferromagnetic Ising model on the pyrochlore lattice, 
where $+$ represents a spin up and $-$ represents a spin down.
It is interesting to note that in the context of an antiferromagnetic
Ising model on the pyrochlore lattice,
 Anderson's model was the second example of what would now
be referred to as a frustrated antiferromagnet, the first one 
being the problem of the Ising antiferromagnet on the triangular lattice studied
by Wannier in 1950 ~\cite{Wannier:1950}.

Anderson's model is an important paradigm for  frustrated 
magnetic systems of interacting spins that reside on the sites of a lattice
of corner-shared triangles or tetrahedra, 
as occurs in kagome, garnet and pyrochlore lattices.
Yet, there are no known realizations of Anderson's Ising antiferromagnet on
the pyrochlore lattice. The reason is that the pyrochlore lattice has a {\it cubic}
symmetry, and there is no energetic reason permitted by symmetry
that would favor a unique global $\hat z$ 
Ising direction to the detriment of another global direction.
However, a system of antiferromagnetically coupled
isotropic Heisenberg spins on the pyrochlore lattice is realistic.
This problem
was the topic of an important 1979 paper~\cite{Villain:1979}. 
In that work, Villain anticipated the failure of the classical Heisenberg
pyrochlore antiferromagnet to develop long range order
down to zero temperature~\cite{Reimers:MC,Moessner:PRB1998} and coined
the name {\it collective paramagnet} to this system $-$ 
a terminology that one may want to ascribe to the classical variant
of the more modern terminology of {\it spin liquid} used in quantum
spin systems.
 Because they lack the intrinsic propensity to develop classical 
long range magnetic order,  
antiferromagnetic materials with Heisenberg spins that reside on 
a pyrochlore lattice are expected to be excellent candidates to
display exotic  quantum mechanical ground states.
It is the anticipation of the observation of interesting magnetic
and thermodynamic behaviors in the broad familly of 
magnetic pyrochlore oxide materials, of generic formula
A$_2$B$_2$O$_7$, that led at the end of the 1980's
to the rapid rise of experimental and theoretical efforts devoted
to the study of these systems~\cite{Greedan:2001,GGG:RMP}.
It is also 
this general scientific endeavour that led to the discovery of  
{\it spin ice}~\cite{Harris:1997,Harris:JMMM1998,Ramirez:Nature1999},
a novel class of frustrated {\it ferromagnet} Ising systems, 
close analogues of both water ice and the cation ordering in spinels.
This discovery also led to the rebirth of a disguised 
variant of Anderson's antiferromagnetic Ising model~\cite{Anderson:1956}
 on the pyrochlore lattice.

\section{Discovery of Spin Ice}

\subsection{Rare earth pyrochlore oxides $-$ generalities}
\label{rare-earth-pyro}

Before reviewing the discovery of spin ice materials and the spin ice problematic,
we first briefly discuss some general aspects of the pyrochlore oxides, the
broader class of materials to which spin ices belong.
We also review the elementary background behind the
origin of the magnetism in the rare earth ions involved in spin ices,
how they acquire their strong Ising nature, and what the
predominant interactions between the magnetic ions are.

The A$_2$B$_2$O$_7$ magnetic pyrochlore oxides form a broad familly
of materials that exhibit a wide range of interesting thermodynamic
and magnetic phenomena, a number of which are still poorly
understood~\cite{Greedan:2001,GGG:RMP}.
Here, both the trivalent A$^{3+}$ ion 
(A is a rare earth element such as Gd, Tb, Dy, Ho, etc, or Y) and the
tetravalent B$^{4+}$ ion (B=Ti, Sn, Mo, Mn) reside on two independent and
interpenetrating pyrochlore sublattices. Figure \ref{pyrochlore} 
shows only one of those two sublattices, say the A sublattice.
 In A$_2$B$_2$O$_7$, one can have either the A or the B
sublattice occupied by a magnetic ion, 
as in Tb$_2$Ti$_2$O$_7$ ~\cite{Gardner:1999} and 
Y$_2$Mo$_2$O$_7$ ~\cite{Greedan:1986}, 
respectively, or both sites can 
 be occupied by a magnetic ion such as in Tb$_2$Mo$_2$O$_7$ ~\cite{Gaulin:1992}.
As discussed by Villain already thirty years ago~\cite{Villain:1979}, 
the pyrochlore lattice is highly frustrated
when the ions carry an isotropic Heisenberg
spin and interacts with nearest neighbors via an antiferromagnetic exchange coupling.
The pyrochlore oxides can be 
metallic (e.g. Nd$_2$Mo$_2$O$_7$, which
displays an interesting anomalous Hall effect~\cite{Taguchi:Nd2Mo2O7})
or they can be insulating, as is
the case for the  A$_2$Ti$_2$O$_7$ and A$_2$Sn$_2$O$_7$ series~\cite{Greedan:2001,GGG:RMP}.
In this chapter, we shall restrict ouselves to the
insulating A$_2$Ti$_2$O$_7$ and A$_2$Sn$_2$O$_7$ series with Dy and Ho 
as magnetic ion A; 
Ti$^{4+}$ and Sn$^{4+}$ are non-magnetic. 
That is, we are
dealing with only one magnetic pyrochlore sublattice, the A sublattice,
in  A$_2$B$_2$O$_7$, which we henceforth simply refer to ``the pyrochlore lattice''.
Neglecting spin-spin interactions~\cite{Molavian:2007}, the Tb$^{3+}$ ion in 
Tb$_2$Ti$_2$O$_7$ and Tb$_2$Sn$_2$O$_7$ would also be described as an Ising spin 
at sufficiently low temperatures~\cite{Rosenkranz:2000,Gingras:2000}.
However, interactions make the strictly Ising model description of these
two materials invalid~\cite{Molavian:2007}.

The pyrochlore lattice can be conveniently described as a face-centered cubic 
(FCC) lattice
with a primitive (``upward'', or ``downward'') tetrahedron basis cell of four sites 
(see Fig.~\ref{pyrochlore}).
The pyrochlore lattice possesses a trigonal (threefold rotational) symmetry with 
respect to any of the four equivalent $\langle 111\rangle$ cubic lattice directions
(i.e. the diagonals of the cubic cell in Fig.~\ref{pyrochlore}).
For each of the four sites in the tetrahedron
unit cell, it will prove  convenient to use
as {\it local} $\hat z_i$ axis of spin quantization
the  specific $[111]$ cube diagonal that
 passes through a given site $i$ and the middle
of the opposite triangular face (see. Fig.~\ref{pyrochlore}).

One should note that
common water ice at atmospheric pressure, ice I$_h$, has a hexagonal
structure while the magnetic pyrochlore lattice has
a cubic symmetry.
Strictly speaking, the Ising pyrochlore problem is equivalent to
cubic ice, $I_{\rm c}$, and not the hexagonal I$_h$ phase. 
However, this  does not modify the 
``ice-rule'' analogy (or mapping) or the close connection 
between the orientation of the magnetic moments in spin ice and the local
proton coordination in water ice.

\subsection{Microscopic Hamiltonian $-$ towards an effective Ising model}
\label{Section:Hmic}

The spin-orbit interaction is large in rare-earth ions and the total angular momentum,
 ${\bf J}={\bf L}+{\bf S}$,
can be taken as a good quantum number. For a given ion, one can  apply Hund's rules
to determine the isolated (vacuum) electronic ground state. For example, Tb has an 
electronic configuration [Xe]4f$^9$6s$^2$ and Tb$^{3+}$ has 
[Xe]4f$^8$ as electronic ground state configuration. Following Hund's rules, one finds that
L=3 and S=3, hence J=6 for Tb$^{3+}$ and the spectroscopic notation for the
ground multiplet is $^7$f$_6$. 
Similarly, one finds J=8 for Ho$_2$Ti$_2$O$_7$ and J=15/2 for Dy$_2$Ti$_2$O$_7$.
Electrostatic and covalent bonding effects, 
which originate from the crystalline environment,
lead to a lifting of the otherwise free ion ground state (2J+1)-fold electronic degeneracy.
This is the so-called crystal field effect.  The theoretical description
of crystal field effects going beyond
the simple Coulomb point charge description 
of the charges surrounding
a rare earth ion, and which includes
covalency effects and  admixing between electronic multiplets, is a rather
technical problem which we shall not discuss here.
In the forthcoming discussion,
 we will merely assume that the single-ion crystal field energy levels
(of the non-interacting ions) have been ``suitably'' determined. 
This can be done, for example,
via experimental spectroscopic methods (e.g. optical spectroscopy, 
inelastic neutron scattering, etc).
Such a spectroscopic approach allows to determine the 
transition frequencies between 
crystal-field levels and their intensities which can then
be described by a  so-called crystal field Hamiltonian, ${\cal H}_{\rm cf}$
(see for example Refs.~\cite{Rosenkranz:2000,Gingras:2000,Mirebeau:2007}). 
We shall return to this point shortly.
Ideally, one would like to make sure that the energy 
levels so-determined are not strongly
``dressed'' (i.e. affected, or renormalized) by inter-ion interactions described by
 ${\cal H}_{\rm int}$. 
This can sometimes be done by considering a highly magnetically diluted variant of the
system of interest~\cite{Gingras:2000}.
This leads us to introduce the Hamiltonian, ${\cal H}$, needed to describe the minimal 
physics at stake:
\begin{equation}
{\cal H} = {\cal H}_{\rm cf}+{\cal H}_{\rm Z}+{\cal H}_{\rm int}	\;.
\label{Htotal}
\end{equation}
Here 
${\cal H}_{\rm cf}$ is the crystal-field Hamiltonian responsible for the lifting
of the degeneracy of the otherwise free single-ion electronic ground state. 
Its energy eigenstates are the above crystal-field energy levels.
As a first approximation, one can express ${\cal H}_{\rm cf}$ in terms of
polynomial functions of the J$_{i,z}$ and 
${\rm J}_{i,\pm}={\rm J}_{i,x}\pm \imath {\rm J}_{i,y}$ components 
of the angular momentum operator ${\bf J}$.
 For the local symmetry of the rare-earth
ions at the A site in A$_2$B$_2$O$_7$, 
one writes ${\cal H}_{\rm cf}$~\cite{Rosenkranz:2000,Gingras:2000,Mirebeau:2007} as
\begin{equation}
{\cal H}_{\rm cf}= \sum_i \sum_{l,m} B_l^m O_l^m({\bf J}_i)	\; ,
\end{equation}
where  $B_l^m$ are the crystal field coefficients and 
$O_l^m({\bf J}_i)$ are the equivalent
crystal field operators. 
Spectroscopic measurements allow to determine the $B_l^m$ via
a fitting procedure of the energy levels with further constraints from the
observed transition intensities~\cite{Rosenkranz:2000,Mirebeau:2007}.
Because of the Wigner-Eckart theorem, $l\le 6$ for L=3 4f  elements.
For example, $O_2^0=3{\rm J}_z^2-{\rm J}({\rm J}+1)$, 
$O_4^0=35{\rm J}_z^4-(30{\rm J}({\rm J}+1)-25){\rm J}_z^2+3{\rm J}^2({\rm J}+1)-6{\rm J}({\rm J}+1))$
and $O_4^{\pm 3}=c_{\pm}[{\rm J}_z,{\rm J}_+^3\pm {\rm J}_-^3]_+$ with
$[A,B]_+ = (AB+BA)/2$, $c_+=1/2$ and $c_-=-\imath/2$
(see Refs.~\cite{Abragam,Ryabov,Rudowicz}).

In Eq.~\ref{Htotal}
${\cal H}_{\rm Z}=-g_{\rm L}\mu_{\rm B}\sum_i {\bf J}_i \cdot \vec B$
is the Zeeman energy describing the interactions of
the rare earth magnetic ions with the  magnetic field $\vec B$;
$g_{\rm L}$ is the Land\'e factor and $\mu_{\rm B}$ is the Bohr magneton.
Finally, ${\cal H}_{\rm int}$ describes the interactions between the ions. 
We consider the following form for ${\cal H}_{\rm int}$:
\begin{equation}
{\cal H}_{\rm int} = 
-\frac{1}{2}\sum_{(i,j)}
 {\cal J}_{ij}\, {\bf J}_i \cdot  {\bf J}_j +
	\left ( \frac{\mu_0}{4\pi} \right)
	\frac{ (g_{\rm L}\mu_{\bf B})^2} { {2 {r_{\rm nn}}^3} }
	\sum_{(i,j)}
	\frac{( {\bf J}_i\cdot  {\bf J}_j - 3  
{\bf J}_i\cdot \hat r_{ij}\hat r_{ij}\cdot {\bf J}_j)}
		{(r_{ij}/r_{\rm nn})^3}	\; ,
\label{Hdip}
\end{equation}
where $\vec r_j-\vec r_i=r_{ij}\hat r_{ij}$ with $\vec r_i$ the position of ion $i$.
Here ${\cal J}_{ij}$ is the microscopic quantum mechanical exchange constant between
ions $i$ and $j$.
We use the convention that ${\cal J}_{ij}<0$ is antiferromagnetic and
${\cal J}_{ij}>0$ is ferromagnetic.
Note that in Eq.~(\ref{Hdip}), the two sums are carried
over all ions $i$ and $j$, hence there is double-counting, and this
is why there is a pre-factor $1/2$ in front of each sum.
The first term ``originates'' from the interactions 
between the ``real'' electronic spins, 
$\vec S_i$ and $\vec S_j$. 
For simplicity, we only consider here an effective isotropic ``exchange'' 
between the total angular momenta ${\rm J}_i$ and 
${\rm J}_j$ with exchange interactions
${\cal J}_{ij}$.  Such description has proven adequate to describe the physics of spin ice 
materials~\cite{Hertog:2000,Bramwell:2001,Ruff:2005,Yavorskii:2008}.
The four distinct types of symmetry-allowed anisotropic exchange
 interactions on the pyrochlore 
lattice are described in Refs.~\cite{Curnoe:2008,McClarty:2008}
The second term is the long range magnetostatic dipole-dipole
interaction with $r_{\rm nn}$ the distance between nearest neighbors.
The pyrochlore oxides have a conventional cubic unit cell of size $a\sim 10\AA$
with 16 ions (i.e. 4 ``upward'' or ``downward' 
primitive tetrahedron basis cells, see Fig.~\ref{pyrochlore})
per conventional cubic unit cell and $r_{\rm nn}=(a\sqrt{2})/4$.
We now discuss the relative energy scales set by 
 ${\cal H}_{\rm cf}$, ${\cal H}_{\rm Z}$ and ${\cal H}_{\rm int}$.

By far, for the (Ho,Dy)$_2$(Ti,Sn)$_2$O$_7$ materials, which we will show later
are spin ices,  ${\cal H}_{\rm cf}$ sets the largest energy scale in the problem. 
In these four materials, the spectrum of ${\cal H}_{\rm cf}$ consists of a ground
state doublet separated by an energy gap, $\Delta$, of 
typically $\Delta \sim 300$~K, to the first excited
state. The Land\'e factor $g_{\rm L}$ of Ho$^{3+}$ and Dy$^{3+}$ is 5/4 and 4/3,
respectively, and $\mu_{\rm B}=9.27 \times 10^{-24}$~J/T $\approx 0.671$~K/T in $H_{\rm Z}$.
In other words, even for a field of 20 T, which is about the
largest field accessible via commercial in-house magnets, the Zeeman energy scale
is of the order of 10 K, hence much smaller than $\Delta$.
Consequently, when 
  calculating the properties of (Ho,Dy)$_2$(Ti,Sn)$_2$O$_7$ at temperatures
$T<~ 10$ K and in magnetic fields less than 20 T or so, on can safely
consider only the magnetic crystal-field ground state doublet of Ho$^{3+}$ and
Dy$^{3+}$ and ignore the susceptibility contributions from the
excited states as well as the van Vleck susceptibility.
The interaction part, ${\cal H}_{\rm int}$, deserves particular attention. 
In insulating magnetic rare earth materials,
the unfilled 4f spin-carrying orbitals of the rare earth ion
are shielded by 5s and 5p orbitals and 
the 4f orbital overlap between the rare earth ions 
and between rare earths and
O$^{2-}$ is therefore small. Consequently, the effective ${\cal J}_{ij}$
exchange coupling between rare earth ions is much 
smaller in insulating rare earth oxides compared to that of 
transition metal oxides. For example, 
a typical value for ${\cal J}_{ij}$ for nearest neighbors, ${\cal J}$,  is
${\cal J} \sim 66$~mK in Dy$_2$Ti$_2$O$_7$ ~\cite{Hertog:2000,Yavorskii:2008}. 
We can also estimate the dipolar energy scale,
${\cal D}\equiv \mu_0 (g_{\rm L} \mu_{\rm B})^2 / (4\pi {r_{\rm nn}}^3)$.
Plugging in a value $r_{\rm nn} \sim 3.5\AA$ and $g_{\rm L}=4/3$, 
we get ${\cal D} \sim 25$~mK ~\cite{Hertog:2000,Yavorskii:2008}.
Comparing ${\cal J}$ and ${\cal D}$ with
$\Delta\sim 300$~K, one sees that the interaction part
${\cal H}_{\rm int}$ of the full Hamiltonian ${\cal H}$
 is very small compared to the crystal-field part, ${\cal H}_{\rm cf}$~\cite{matrix}.
In practice, this means that 
 the independent/non-interacting single-ion crystal-field
 eigenstates of ${\cal H}_{\rm cf}$ 
form a very convenient basis to describe the Hilbert space of
interacting Dy$^{3+}$ and Ho$^{3+}$ ions in
(Ho,Dy)$_2$(Ti,Sn)$_2$O$_7$.
In particular, since the scale
of  ${\cal H}_{\rm int}$ is so small compared to $\Delta$, one can, for
all practical purposes, neglect all the excited states of ${\cal H}_{\rm cf}$
and work with a reduced Hilbert space solely spanned by the 
two degenerate states of the ground doublet for each magnetic sites.
In such a problem, where the high-energy sector of the theory is
so well separated from the low-energy sector, it is convenient to
introduce a so-called {\it effective Hamiltonian}, ${\cal H}_{\rm eff}$,
that operates only within the low-energy sector. 
The Chapter by Mila and  Schmidt in this book
describes the general methodology as to how 
to derive a ${\cal H}_{\rm eff}$ for a given interacting system where there are
well separated ``reference'' energy sectors.
For the spin ices, the lowest order of the perturbation theory which
defines ${\cal H}_{\rm eff}$ is amply sufficient; i.e.
\begin{eqnarray}
{\cal H}_{\rm eff} & \approx & P{\cal H}_{\rm int}P
\label{PHP}
\end{eqnarray}
with $P$, the projector in the low-energy sector, defined as 
\begin{equation}
P = \sum_{\{k\}} \vert \Phi_{0,\{k\}}\rangle\langle \Phi_{0,\{k\}}\vert
\end{equation}
where
\begin{equation}
\vert \Phi_{0,\{k\}}\rangle
=\prod_{i=1}^N \vert \phi_{i,0}^{(k_i)} \rangle	\; .
\end{equation}
Here $\phi_{i,0}^{(k_i)}$
is the $k_i^{\rm th}$ state of the single-ion crystal-field 
ground doublet of rare earth ion $i$.
Since ${\cal H}_{\rm int}$ is a pairwise Hamiltonian, ${\cal H}_{\rm eff}$ is,
to lowest order and given by Eq.~(\ref{PHP}),
 also a pairwise effective Hamiltonian. One can
then determine ${\cal H}_{\rm eff}$ by considering two ions each in one
of the two states, 
$\vert \phi_{i,0}^{(k_i=+)}\rangle$ and
$\vert \phi_{i,0}^{(k_i=-)}\rangle$, 
of their respective crystal-field ground doublet.
Hence, the effective Hamiltonian ${\cal H}_{\rm eff}$ 
is a $4\times 4$ matrix. At this point, we need to return to 
the crystal-field problem and discuss the nature of the non-interacting
$\vert \phi_{i,0}^{(k_i=\pm)}\rangle $ states
that make each of the single-ion ground doublet.

From inelastic neutron measurements, it is found that 
$\vert \phi_{i,0}^{(k_i=+)}\rangle $ and
$\vert \phi_{i,0}^{(k_i=-)}\rangle $ 
for Ho$_2$Ti$_2$O$_7$ are 
$\vert \phi_{i,0}^{(k_i=\pm)}\rangle  \approx 
\vert {\rm J}=8,m_{\rm J}=\pm 8\rangle$, with
negligible contributions from other 
$\vert {\rm J}=8,m_{\rm J}\rangle$ components.
By rescaling the crystal field parameters determined
for  Ho$_2$Ti$_2$O$_7$ one can determine the crystal field parameters
for Dy$_2$Ti$_2$O$_7$~\cite{Rosenkranz:2000}. One finds for the latter material
$\vert \phi_{i,0}^{(k_i=\pm)}\rangle \approx \vert 
{\rm J}=15/2,m_{\rm J}=\pm 15/2\rangle$ 
with again negligible weight 
from other $\vert {\rm J}=15/2,m_{\rm J}\rangle$ components~\cite{Rosenkranz:2000}.
 It is important to recall the above discussion
about the choice of the $\hat z_i$ quantization direction.
Here, the magnetic quantum number $m_{\rm J}$ eigenstate 
of ${\rm J}_z$ refers to the component of ${\bf J}_i$ along the 
{\it local}  $\hat z_i$ direction that points along one of the four
cubic $\langle 111\rangle$ directions.  
Calculating the matrix elements of $P{\cal H}_{\rm int}P$, one finds that,
 for both  Ho$_2$Ti$_2$O$_7$ and Dy$_2$Ti$_2$O$_7$,
the only significant matrix elements are  
$\langle \phi_{i,0}^{(k_i=\pm)} 
  \vert {\rm J}_z \vert \phi_{i,0}^{(k_i=\pm)} \rangle \approx \pm {\rm J} $.
The physical meaning is simple. In both
Ho$_2$Ti$_2$O$_7$ and  Dy$_2$Ti$_2$O$_7$, the magnetic moment
can, for all practical purposes,
only point parallel or antiparallel to the local $[111]$ direction;
we are deadling with a local $\langle 111 \rangle$ Ising model!

It is useful to write down the 
set of local $\hat z_i$ quantization axis  
(i.e. $\hat z_i \parallel \langle 111 \rangle$ axis). 
We take 
$\hat z_1=\frac{1}{\sqrt 3}( \hat x+\hat y+\hat z)$,
$\hat z_2=\frac{1}{\sqrt 3}(-\hat x-\hat y+\hat z)$,
$\hat z_3=\frac{1}{\sqrt 3}(-\hat x+\hat y-\hat z)$ and 
$\hat z_4=\frac{1}{\sqrt 3}( \hat x-\hat y-\hat z)$, with
$\hat z_\mu \cdot \hat z_\nu = -1/3$ for $\mu\ne \nu$.
To proceed formally, one could 
re-express the above $4\times 4$ matrix representing 
${\cal H}_{\rm eff}$  in terms of tensor products of
Pauli matrices, $\sigma_i^{\alpha_i}\otimes \sigma_j^{\beta_j}$ where we are
here using a {\it local} orthogonal $\{x_i,y_i,z_i\}$ frame~\cite{Molavian:2007},
where $\alpha_i=x_i,y_i,z_i$.
Had we kept the full $\vert {\rm J},m_{\rm J}\rangle$ decomposition of
$\vert \phi_{i,0}^{(k_i=\pm)} \rangle$,
one would have found  that 
the coefficient of the $\sigma_{i}^{z_i}\sigma_{j}^{z_j}$ term 
has by far the largest coefficient in ${\cal H}_{\rm eff}$.
We thus end up with a classical model
with Ising spins pointing parallel or
antiparallel to their {\it local} $[111]$ direction.

Alternatively, one could have proceeded much 
more straightforwardly and simply replace 
\begin{equation}
{\bf J}_i \rightarrow \vert \langle {\rm J}_z\rangle\vert \sigma_i^z \hat z_i 
\end{equation}
right at the outset, and use 
$\vert \langle {\rm J}_z\rangle \vert \approx 8$ for Ho$_2$Ti$_2$O$_7$ 
and  $\vert \langle {\rm J}_z\rangle \vert \approx 15/2$ for Dy$_2$Ti$_2$O$_7$.
Indeed, this is what was implicitely done in most previous
numerical studies of dipolar spin
 ice~\cite{Ramirez:Nature1999,Hertog:2000,Bramwell:2001,Ruff:2005,Siddharthan:1999,butnot}.
Since there are no other components $\sigma_i^{\alpha_i}$ with $\alpha_i\ne z_i$
that do not commute among themselves
left in ${\cal H}_{\rm eff}$,  
we have a strictly classical Ising model for $H_{\rm eff}$. One can therefore 
work in the basis of eigenstates of $\sigma_i^{z_i}$
and, henceforth, we merely treat $\sigma_i^{z_i}=\pm 1$ as a classical variable.
Admitedly, 
the discussion here leading to ${\cal H}_{\rm eff}$ as a classical 
Ising model for the Dy-based and Ho-based oxyde pyrochlores is rather academic.
However, as mentioned earlier in Section~\ref{rare-earth-pyro}
 it is not so for Tb$_2$Ti$_2$O$_7$~\cite{Gardner:1999}
 and Tb$_2$Sn$_2$O$_7$~\cite{Mirebeau:2005}.
Indeed, for these two materials, the energy gap separating the ground doublet
and the excited doublet is not so large compared to the exchange
and dipole-dipole interactions. In that case, higher
order terms in the perturbation expansion leading to ${\cal H}_{\rm eff}$
must be retained
(again, see 
the chapter by Mila and  Schmidt) in order
to derive a proper quantum theory of this material, as was recently done 
for a simple model of Tb$_2$Ti$_2$O$_7$~\cite{Molavian:2007}.

Returning to ${\cal H}_{\rm int}$ above, and writing
${\rm J}_i \rightarrow \vert \langle
 {\rm J}_z\rangle\vert \sigma_i^z \hat z_i$, we 
arrive at our effective {\it classical} $\langle 111\rangle$ pyrochlore Ising model
\begin{equation}
{\cal H}_{\rm DSM} = 
-\frac{1}{2}\sum_{(i,j)}
 J_{ij}\, (\hat z_i \cdot \hat z_j)\,  \sigma_i^{z_i} \sigma_j^{z_j} \; + \;
	\frac{D}{2} 	\sum_{(i,j)}
	\frac{(\hat z_i\cdot \hat z_j - 3 \hat z_i\cdot \hat r_{ij}\hat r_{ij}\cdot\hat z_j)}
		{(r_{ij}/r_{\rm nn})^3}	\, \sigma_i^{z_i}\sigma_j^{z_j} \; , 
\label{Hdsm}
\end{equation}
which we shall henceforth refer to as the ``dipolar spin ice model'' (DSM).
Here we have written $J_{ij}={\cal J}_{ij}\langle {\rm J}_z\rangle ^2$
and
$D=\mu_0 (g_{\rm L}\mu_{\rm B}\langle {\rm J}_z
\rangle)^2/(4\pi {r_{\rm nn}}^3)$.
Note that  the variables $\sigma_i^z=\pm 1$ have become
simple labels indicating whether 
the $z_i$-component of ${\bf J}_i$ points ``in''
or ``out'' of a primitive tetrahedron unit cell characterized by the above
set $\{\hat z_1,\hat z_2,\hat z_3,\hat z_4\}$
(see Fig.~\ref{pyrochlore}).
Having established that the magnetic Dy$^{3+}$ and Ho$^{3+}$ ions
in  Dy$_2$Ti$_2$O$_7$ and 
Ho$_2$Ti$_2$O$_7$ should be described by effective classical Ising spins
at temperatures much lower than the lowest crystal field gap $\Delta$, we can 
now proceed to discuss the experimental behavior of these two materials, and how
the spin ice phenomenology arises.

\subsection{Discovery of spin ice in Ho$_2$Ti$_2$O$_7$}

In a 1997 paper, Harris, Bramwell and collaborators reported the results of
a neutron scattering study of Ho$_2$Ti$_2$O$_7$~\cite{Harris:1997}.
They found that, in zero applied magnetic field, no evidence of a transition
to long range order could be detected down to 0.35 K via neutron scattering,
while muon spin relaxation results could rule out a transition down 
to 0.05 K~\cite{Harris:JMMM1998}.
The most surprising part of those results 
was that the Curie-Weiss temperature, $\theta_{\rm CW}$,
was found to be {\it positive} with $\theta_{\rm CW} \approx +1.9$~K, hence indicating
overall ferromagnetic interactions. Naively, one would have expected such a 
three-dimensional cubic system with ferromagnetic interactions to develop long range 
order at a critical temperature of the same order of magnitude as $\theta_{\rm CW}$.
It was also found that the magnetic field dependence of the
neutron scattering intensity depends on the protocol followed to
magnetize the sample~\cite{Fennell:2005}. 
In other words, the system displayed a history 
dependence reminiscent of what is 
observed in random spin glasses~\cite{Binder:1986}, 
although
no significant random disorder is present in Ho$_2$Ti$_2$O$_7$.

The authors of Refs.~\cite{Harris:1997,Harris:JMMM1998} proposed that the strong local
$[111]$ Ising anisotropy of Ho$^{3+}$ in Ho$_2$Ti$_2$O$_7$
frustrates the development of ferromagnetic order.
They considered a simple model of ferromagnetically 
coupled $\langle 111\rangle$ Ising spins 
on the pyrochlore lattice~\cite{Bramwell:JPC1998}.
By doing so, they
established the connection between their model
and that of Pauling's model for the problem of proton disorder in water ice,
hence coining the name {\it spin ice model} to their model (see Fig.~\ref{spinice}).
For clarity sake, and since we shall discuss below the important
role that long range dipole-dipole interactions play in spin ice materials, we
relabel Harris {\it et al.}'s model as {\it nearest-neighbor spin ice model},
in order to distinguish it from the {\it dipolar spin ice model} (DSM) of
Eq.~(\ref{Hdsm}).
We discuss in the next section this nearest-neighbor spin ice model and show
that it has the same residual entropy as water ice.

\subsection{Nearest-neighbor ferromagnetic $\langle 111\rangle$ 
Ising model and Pauling's entropy}

\label{Section:NNSI}

\subsubsection{Nearest-neighbor spin ice model}

Consider a simplified version of the Ising Hamiltonian in Eq.~(\ref{Hdsm}) 
where the dipolar interaction coefficient is
set to $D=0$ for the time being and where
we restrict the exchange interactions, $J_{ij}$, solely to nearest neighbors.
We thus have
\begin{equation}
{\cal H}_{\rm nn}= -J \sum_{\langle i,j\rangle} 
(\hat z_i\cdot \hat z_j)\, \sigma_i^{z_i} \sigma_j^{z_j}
\label{Hnn}
\end{equation}
with the convention that the nearest-neighbor exchange $J>0$ is ferromagnetic 
while $J<0$ is antiferromagnetic.
Taking $\hat z_i \cdot \hat z_j=-1/3$ for nearest neighbors on the pyrochlore, 
we have:
\begin{eqnarray}
{\cal H}_{\rm nn}
& = & -J \sum_{\langle i,j\rangle}(\hat z_i\cdot \hat z_j)\, \sigma_i^{z_i} \sigma_j^{z_j} \nonumber \\
& = & -J (-\frac{1}{3})
  \sum_{\langle i,j\rangle}\sigma_i^{z_i} \sigma_j^{z_j} \nonumber \\
& = &  \frac{J}{3} \sum_{\langle i,j\rangle}\sigma_i^{z_i} \sigma_j^{z_j} 		\; .
\label{Hnnsi}
\end{eqnarray}
We can  rewrite Eq.~(\ref{Hnnsi}) as ~\cite{Moessner:PRB1998}
\begin{eqnarray}
{\cal H}_{\rm nn}
& = & \frac{J}{6} \sum_{\Delta} (L_\Delta)^2 - \frac{2N_\Delta J}{3}
\label{Hnnsi-teta}
\end{eqnarray}
where the $\sum_\Delta$ is now carried over each ``upward'' and ``downward'' tetrahedron
units 
labelled by $\Delta$, with a total of $N_\Delta$ ``upward'' and ``downward'' tetrahedra,
and where $L_\Delta$ is the total spin on unit $\Delta$, defined as
\begin{eqnarray}
L_\Delta \equiv \sigma_{\Delta}^{z_1}
+\sigma_{\Delta}^{z_2}+\sigma_{\Delta}^{z_3}+\sigma_{\Delta}^{z_4}	\; .
\label{Deltasum}
\end{eqnarray}

From Eq.~(\ref{Hnnsi}), one can appreciate the ``first magic'' of spin ice.
We will discuss the ``second magic'' of spin ice in Section~\ref{Section:competing-int}.
By starting with a ferromagnetic nearest-neighbor exchange, 
$J>0$ in Eq.~(\ref{Hnn}),  one ends  up with 
$\sigma_i^{z_i}$ Ising variables that 
are now coupled via an effective {\it antiferromagnetic} coupling constant in Eq.~(\ref{Hnnsi}).
This is the same model as 
Anderson's frustrated antiferromagnetic pyrochlore Ising model used
to describe the problem of cation ordering in inverse spinels~\cite{Anderson:1956}. 
The ferromagnetic $\langle 111\rangle$ 
pyrochlore Ising model is therefore frustrated and,
just as Anderson's model, it must possess a residual Pauling entropy.

In terms of ${\cal H}_{\rm nn}$ in Eq.~\ref{Hnnsi-teta}, the ground state 
configuration on a single tetrahedron for antiferromagnetic $J<0$
consists of all $\sigma_i^{z_i}=+1$ or all $\sigma_i^{z_i}=-1$.
In terms of the ``real''  ${\bf J}_i$ spins in Eq.~\ref{Hdip}, 
this corresponds to all ${\bf J}_i$ pointing ``in'' 
or all ${\bf J}_i$ 
pointing ``out'' of a reference tetrahedron unit cell. Globally, there are only two
such states and, upon cooling the system, a second order transtion to a four
sublattice antiferromagnetic
N\'eel ordered state in the three-dimension Ising universality class occurs.
The four sublattice N\'eel ordered phase described in terms of
the ${\bf J}_i$ maps onto a ferromagnetic ground state in terms of the $\sigma_i^{z_i}$.
The pyrochlore lattice with antiferromagnetically coupled $\langle 111\rangle$ spins
is therefore {\it not} frustrated.
Conversely, for ferromagnetic $J>0$, the 
$\sigma_i^{z_i}$ Ising variables are effectively coupled {\it antiferromagnetically}
and the ground state on a single tetrahedron is 6-fold degenerate, 
arising from the condition
$\sigma_{\Delta}^{z_1}+\sigma_{\Delta}^{z_2}+\sigma_{\Delta}^{z_3}+\sigma_{\Delta}^{z_4}=0$
on each tetraderon $\Delta$, from Eq.~(\ref{Deltasum}),  i.e, two positive
and two negative $\sigma_i^{z_i}$. 
This is the incipient spin
configuration leading to the 
extensive Pauling entropy 
of spin ice in the thermodynamic limit~\cite{Ramirez:Nature1999}.

We now discuss how the nearest-neighbor spin ice model possesses a Pauling entropy. 
We invite the reader to refer to Fig.~\ref{spinice} where a more direct physical, 
or geometrical, connection beween the $\sigma_i^{z_i}$ configurations 
in the spin ice problem and that of the ``real'' proton positions in water ice
is illustrated.

\subsubsection{Pauling entropy of the nearest-neighbor spin ice model}

The macroscopically degenerate ground states that are often 
characteristic of frustrated systems can be understood in terms of
an underconstraint argument~\cite{Moessner:PRB1998}. 
That is, the extensive degeneracy arises from 
the difference between the number of 
constraints necessary to determine a ground state 
and the number of degrees of freedom that the system possesses. 
By following Pauling's heuristic argument for water ice, one can 
calculate the residual entropy of spin ice. 
Consider Anderson's Ising pyrochlore antiferromagnet, 
to which the local $\langle 111\rangle$ pyrochlore Ising model
maps,
as discussed in Section~\ref{cation-ordering}. 
The  ground state condition is ``underconstrained'',
demanding only that the total magnetization, $L_\Delta$, of the four Ising spins on
each tetrahedron fulfills $L_\Delta=0$.
Six of the $2^4=16$ possible spin configurations satisfy this condition. 
Counting $2^4$ configurations for each tetrahedron gives, 
for a system of $N$ spins and $N/2$
``upward'' and ``downward'' tetrahedra,  $(2^4)^{N/2}=4^{N}$  microstates.
This number drastically overestimates the exact total of $2^N$ microstates
for $N$ Ising spins.
The reason is that each spin is shared between two
tetrahedra, hence the above 16 configurations on each tetrahedron are not independent. 
Borrowing Pauling's argument, we allocate $2^2$ states
per tetrahedron and, assuming that $6/16$ of them satisfy the constraint,
we obtain a ground state degeneracy
$\Omega_0=\{(2^2)^{N/2}(6/16)\}^{N/2}=(3/2)^{N/2}$. 
The corresponding entropy,
$S_0=k_{\rm B}\ln(\Omega_0)=(Nk_{\rm B})/2\ln(3/2)$, 
is of course just Pauling's 
original result revamped for the pyrochlore spin ice problem. 
Another way to obtain the residual entropy is as follows.
For a pyrochlore lattice with $N$ spins, there are 
$N/4$ tetrahedron unit cells.
Take all the ``upward'' tetrahedron unit cells in Fig.~\ref{pyrochlore}.
If all these tetrahedra are in a spin ice state, there
can be $6^{N/4}$ independent spin ice configurations in the system. 
However, not all these states are valid ground states. 
The $N/4$  ``downward'' tetrahedra, which
are formed by the corners of these "upward" tetrahedra,
must also satisfy the ice rules.
The probability that a random tetrahedron satisfies the ice rule is 
6/16; so the above number should be deflated
by this scale factor for each ``downward'' tetrahedron, hence giving
a total of $6^{N/4}\times (6/16)^{N/4}=(3/2)^{N/2}$ ice-rule obeying states
for the whole system, as obtained above.
Finally, one could have also simply directely borrowed Pauling's result.
In the mapping from water ice to Anderson's antiferromagnet, or
alternatively to spin ice, $N$ oxygen O$^{2-}$ ions 
correspond to $2N$ spins.
We shall now show that this Pauling's estimate,  $S_0=(Nk_{\rm B}/2)\ln(3/2)$, 
for spin ices, is in good agreement with experimental results.

\subsection{Residual entropy of Dy$_2$Ti$_2$O$_7$ and Ho$_2$Ti$_2$O$_7$}

\begin{figure}[ht]
\begin{center}
\includegraphics[height=8cm]{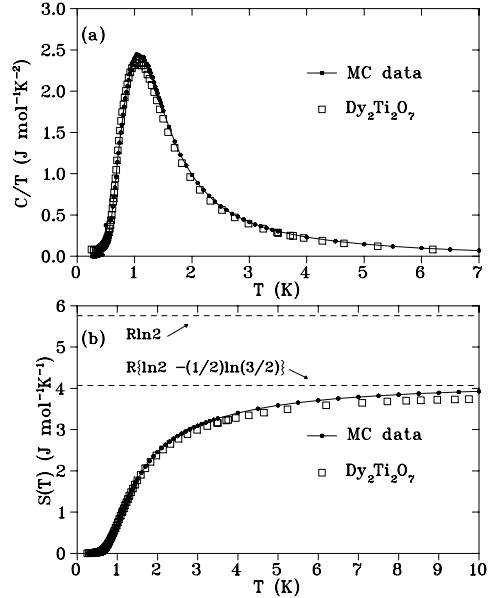}
\caption{(a) Specific heat and (b) entropy data for
Dy$_{\rm 2}$Ti$_{\rm 2}$O$_{\rm 7}$
from Ref.~\cite{Ramirez:Nature1999}, 
compared with Monte Carlo simulation results~\cite{Hertog:2000}
for the dipolar spin ice model, with $J/3=-1.24$~K and $5D/3=2.35$~K.
Two temperature regimes can be identified.
(i) At a temperature $T$ much higher than the peak temperature, $T_{\rm peak}\sim 1.24$ K,
the system is in the paramagnetic regime and is weakly correlated and
individual tetrahedra do not obey the ``two-in/two-out'' ice rules.
As the temperature approaches $T_{\rm peak}$, the ice-rules become progressively fulfilled.
The Schottky-like peak in $C$ arises 
when the temperature drops below the energy gap between the ice-rule obeying states 
and the excited ``three-in/one-out'' states 
and ``all-in/all-out'' states.
(ii) As $T$ drops below $T_{\rm peak}$, 
the spin flip rate drops exponentially rapidly~\cite{Melko:2004}
as the system settles in an ice-rule obeying state with two spins ``in'' and two spins ``out'' on
each tetrahedron. There is no phase transition between the high temperature paramagnetic state
($T>T_{\rm peak})$ and the spin ice regime at $T<T_{\rm peak}$. The spin ice regime can 
therefore be described as a collective paramagnet~\cite{Villain:1979}.
(Figure reprinted with permission from
B. ~C. den Hertog and M. ~J. ~P. Gingras,
Phys. Rev. Lett. \textbf{84}, 3430 (2000).
Copyright 2000 by the American Physical Society.)
}
\label{Dy2Ti2O7-Cv}
\end{center}
\end{figure}

The first compelling thermodynamic evidence for the existence of a spin
ice state in Ising pyrochlore systems was obtained via measurements of
the magnetic specific heat, $C(T)$, on Dy$_2$Ti$_2$O$_7$~\cite{Ramirez:Nature1999}.
This material with magnetic Dy$^{3+}$ is, as Ho$_2$Ti$_2$O$_7$, 
characterized by a strongly Ising-like ground
state doublet that is separated by a large energy of 
approximately 300 K from the excited doublet. 
The temperature
dependence of $C(T)$ is shown in the upper panel of Fig.~\ref{Dy2Ti2O7-Cv}.
To determine the residual magnetic entropy of Dy$_2$Ti$_2$O$_7$, 
Ramirez and co-workers followed 
an approach similar to the one that Giauque and colleagues used
to determine the entropy of water ice~\cite{Giauque:1933,Giauque:1936}.

In general, one can only measure the
change of entropy between two temperatures. Giauque 
and collaborators
calculated
the entropy change of water between 10 K 
 and the gas
phase of H$_2$O by integrating the specific heat divided by temperature, 
$C(T)/T$, adding to it the latent heat at the melting and vaporization transitions.
They then compared this value with the {\it absolute} value expected 
for the entropy calculated for an ideal (non-interacting)  
gas phase using inputs from spectroscopic measurements which allowed
to determine the rotational and vibrational energy spectrum,
hence the roto-vibrational entropy, and adding to it the
translational entropy of an ideal gas given by the Sackur-Tetrode equation.
The difference between the measured value and the
theoretically expected value gave the residual entropy of water ice,
whose origin was explained by Pauling~\cite{Pauling:1935}.

Ramirez {\it et al.} measured the magnetic specific heat of
a powder sample of Dy$_2$Ti$_2$O$_7$ between $T_1$=300 mK, ``deep'' inside
the frozen ice regime, and $T_2$=10 K, in the
paramagnetic regime, where the expected entropy per mole 
should be $R\ln(2)$ for a two-state system
(${R}=N_0 k_{\rm B}$ is the gas molar 
constant and $N_0$ is the Avogadro number).

The entropy change between $T_1$ and $T_2$, $\Delta S_{1,2}$, was found by
integrating $C(T)/T$ between these two temperatures:
\begin{equation}
\Delta S_{1,2} = S(T_2)-S(T_1) = \int_{T_1}^{T_2} \frac{C(T)}{T} \, dT.
\end{equation}

The lower panel of Fig.~\ref{Dy2Ti2O7-Cv} 
shows that the magnetic entropy recovered is approximately 3.9
Jmol$^{-1}$K$^{-1}$, a value quite a bit smaller than 
R$\ln(2)\approx$ 5.76 Jmol$^{-1}$K$^{- 1}$.
The difference, 1.86 Jmol$^{-1}$K$^{- 1}$ is fairly 
close to the above Pauling-like estimate for the
entropy associated with the extensive
degeneracy of ice: $(R/2)\ln(3/2) = 1.68$ Jmol$^{-1}$K$^{-1}$, consistent
with the existence of an ice-rule obeying
spin ice state in Dy$_2$Ti$_2$O$_7$.
More recent measurements~\cite{Ke:2007} on single crystals of Dy$_2$Ti$_2$O$_7$ have
found a specific heat below 1.5 K that differs quantitatively fairly
significantly from that of Ref.~\cite{Ramirez:Nature1999},
ultimately giving a much better agreement between the experimentally determined
residual entropy of the material and Pauling's value.

As discussed above, Ho$_2$Ti$_2$O$_7$ was the first material proposed as a
spin ice system. It turns out that Ho$_2$Ti$_2$O$_7$ is less
convenient to perform low temperature specific heat measurements than 
Dy$_2$Ti$_2$O$_7$. The origin of this difficulty stems from the
unusually large hyperfine interaction between the nuclear and electronic
spins of Ho$^{3+}$. This interaction leads to a Schottky anomaly in the specific
heat at a temperature $T\sim 0.3$~K which  significantly obscures the otherwise
purely electronic broad specific heat feature arising from the formation of
the ice-rule obeying low-energy manifold~\cite{Hertog:condmat}.
Once the nuclear contribution to the specific heat has been subtracted, the 
exposed electronic contribution and the consequential residual Pauling
entropy of the spin ice state in Ho$_2$Ti$_2$O$_7$ 
can be revealed~\cite{Bramwell:2001,Cornelius:2001}.

In summary, the Ho$_2$Ti$_2$O$_7$ and Dy$_2$Ti$_2$O$_7$ possess a residual
low-temperature entropy compatible with that estimated on the basis of  a
Pauling argument applied to the ferromagnetic
nearest-neighbor $\langle 111 \rangle$ Ising model
on the pyrochlore lattice.  
The related Ho$_2$Sn$_2$O$_7$~\cite{Kadowaki:2002} and
Dy$_2$Sn$_2$O$_7$~\cite{Ke:0707.4009} are also spin ice systems.

\section{Dipolar Spin Ice Model}

\subsection{Competing interactions in dipolar spin ice model}
\label{Section:competing-int}

In Section~\ref{Section:NNSI}, 
we invoked a simple nearest-neighbor ferromagnetic
$\langle 111 \rangle$ Ising model 
to rationalize the spin ice phenomenology 
in real materials.
 Yet, it was argued in Section~\ref{Section:Hmic} that magnetic dipole-dipole
interactions are often of sizeable strength in rare-earth magnetic systems.
In the case of Dy$_2$Ti$_2$O$_7$ and Ho$_2$Ti$_2$O$_7$, the value of 
$D=\mu_0(g_{\rm L} \mu_{\rm B} \langle {\rm J}_z \rangle )^2/(4\pi {r_{\rm nn}}^3)$ 
is estimated at $D\sim 1.4$~K~\cite{Hertog:2000,Siddharthan:1999,butnot},
 which is comparable to the experimentally
measured Curie-Weiss temperature $\theta_{\rm CW}$ in these materials.
Even if one was assuming that the nearest-neighbor exchange 
$J \sim \theta_{\rm CW}\sim 1$~K, one is still in a regime where 
the dipolar interactions are comparable in magnitude to 
the nearest-neighbor exchange 
interactions~\cite{Ramirez:Nature1999,Hertog:2000,Siddharthan:1999}.
This observation raises a paradox.
The existence of a ground state with extensive degeneracy should in
principle results from the
underconstraints that the Hamiltonian imposes on the spin configurations that
minimize the classical ground state energy~\cite{Moessner:PRB1998}. 
Consider the dipolar interactions in
Eq.~(\ref{Hdsm}). 
First of all, they are rather complicated as they couple spin and space 
directions via the $(\hat z_i\cdot \hat r_{ij})(\hat r_{ij}\cdot \hat z_j)$ term.
Secondly, dipolar interactions are very long-ranged, decaying as $1/{r_{ij}}^3$ with
distance $r_{ij}$ separating ions $i$ and $j$. How can one thererefore understand
that the dipolar interactions allow for the emergence of an extensively
degenerate spin ice state at a temperature $T$ of order of $D$
with spin couplings as ``complicated'' (i.e. anisotropic and long-range) as the
dipolar interactions?
How dipolar interactions manage to do so is the ``second magic'' of dipolar spin ice.

One way to first investigate this question is to 
go beyond the nearest-neighbor ferromagnetic $\langle 111 \rangle$
Ising model and perform Monte Carlo simulations 
where the dipolar interactions of Eq.~(\ref{Hdsm}) are taken into 
account~\cite{Ramirez:Nature1999,Hertog:2000,Bramwell:2001,Siddharthan:1999,Ruff:2005,Yavorskii:2008,Melko:2001,Melko:2004}.
The first studies of this problem considered truncated dipolar
interactions beyond a certain cut-off 
distance~\cite{Ramirez:Nature1999,Siddharthan:1999}.
It was then followed by another work~\cite{Hertog:2000}
that incorporated the true long-range dipolar interactions using
the so-called Ewald summation method~\cite{Melko:2004}  which is commonly
used to handle long-range Coulomb and dipolar interactions~\cite{Kittel}.

\begin{figure}[ht]
\begin{center}
\includegraphics[height=14cm,width=7cm]{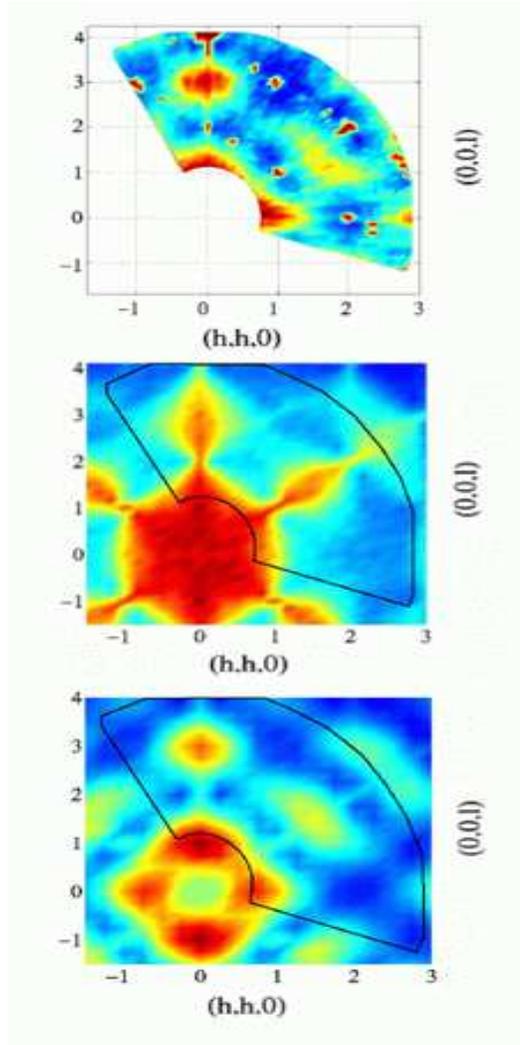}
\caption{ ${\rm Ho_2Ti_2O_7}$: Neutron scattering in the $(hhl)$
plane showing experimental data (upper panel; the sharp spots are
nuclear Bragg scattering with no magnetic component), compared
with Monte Carlo simulations of the near neighbour spin ice model
middle panel) and dipolar model (lower panel)~\cite{Bramwell:2001}.
Blue (light) indicates the weakest and red-brown (dark) 
the strongest intensity.
(Figure reprinted with permission from
S. ~T. Bramwell {\it et al.},
Phys. Rev. Lett. \textbf{87}, 047205 (2001).
Copyright 2001 by the American Physical Society.)
}
\label{Ho2Ti2O7-neutrons}
\end{center}
\end{figure}

Consider
 ${\cal H}_{\rm DSM}$ in Eq.~(\ref{Hdsm}) with long-range dipole-dipole interactions,
$D\ne 0$, and only nearest-neighbor exchange $J_{ij}=J$ as a starting model to describe 
Dy$_2$Ti$_2$O$_7$.
One can take the dipolar inteaction coupling $D$ as a more
or less a priori known quantity if one makes the reasonably accurate
approximation that the crystal field ground state doublet is essentially
composed of only the $\vert {\rm J}=15/2,
m_{\rm J}=\pm 15/2\rangle$ components.
This gives $D\sim 1.4$~K ~\cite{Hertog:2000}.
Hence, the nearest-neighbor exchange $J$ is the only
unknown parameter in the model.
In the Monte Carlo simulations~\cite{Hertog:2000}, 
it was found that fitting either the
height of the specific peak {\it or} the temperature at which the
peak occurs allows for a unique and consistent determination of $J$.
The values $J/3 \approx -1.24$~K~\cite{Hertog:2000,JandJnn}
and $J/3 \approx -0.55$~K~\cite{Bramwell:2001} were
estimated for
 Dy$_2$Ti$_2$O$_7$~\cite{Hertog:2000} and
 Ho$_2$Ti$_2$O$_7$~\cite{Bramwell:2001}, respectively.
Interestingly, the nearest-neighbor interaction
$J$ is {\it antiferromagnetic} in both cases which should, 
on its own, and according to
the discussion in Section~\ref{Section:NNSI},
give rise to a four sublattice long-range N\'eel ordered phase~\cite{Hertog:2000}.
This already indicates that it must 
be the dipolar interactions that are responsible for the spin ice phenomenology.
Figure~\ref{Dy2Ti2O7-Cv} 
shows that a Monte Carlo simulation of the dipolar spin ice model
(DSM) in Eq.~(\ref{Hdsm}), 
with $J/3 \approx -1.24$~K and $5 D/3 \approx 2.35$~K, 
gives an excellent description of the magnetic
specific heat data $C(T)$ of Dy$_2$Ti$_2$O$_7$. Also, as a consequence, the
integration of the Monte Carlo $C(T)/T$  reproduces
the experimental Pauling-like residual entropy (see 
lower panel of Fig.~\ref{Dy2Ti2O7-Cv}).
A similar agreement 
between experimental and Monte Carlo data for the
specific heat was found for Ho$_2$Ti$_2$O$_7$ using 
with $J/3 \approx -0.55$~K and $5D/3  \approx 2.35$~K~\cite{Bramwell:2001}.
Furthermore,
 in Ho$_2$Ti$_2$O$_7$, it was possible to perform neutron scattering studies 
on a single crystal~\cite{Bramwell:2001}. After having fixed 
$J$ from $C(T)$ data on Ho$_2$Ti$_2$O$_7$, 
it was found that the scattering wave vector ${\vec q}$ dependence
of the experimental neutron scattering intensity was well reproduced
by results from Monte Carlo simulations~\cite{Bramwell:2001}, 
as illustrated in Fig.~\ref{Ho2Ti2O7-neutrons}.

\newpage

\newpage

Hence, despite their complex structure, not only are dipolar
interactions compatible with the existence of a degenerate state, but in 
fact they appear to {\it be responsible} for it, 
since the nearest-neighbor exchange interaction $J$, being antiferromagnetic in
Ho$_2$Ti$_2$O$_7$ and Dy$_2$Ti$_2$O$_7$, 
would by itself lead to long range order.
Hence, where does the spin ice phenomenology come 
from in the dipolar spin ice model of Eq.~(\ref{Hdsm}) with
long range dipole-dipole interactions?
A first way to address this question is
to take the dipolar interactions and truncate them at nearest-neighbor
distance. On the pyrochlore lattice, we have
$\hat z_i\cdot \hat z_j=-1/3$ and 
$(\hat z_i \cdot \hat r_{ij}) (\hat r_{ij}\cdot z_j)=-2/3$ for nearest
 neighbor ions $i$ and $j$. We thus have
\begin{equation}
H = \sum_{\langle i,j\rangle} \left (\frac{J}{3}+\frac{5D}{3} 	\right ) \,
\sigma_i^{z_i}\sigma_j^{z_j} + H^{\rm dip}_{>r_{\rm nn}} 
\label{SIcriterion}
\end{equation}
where we have incorporated all dipolar interactions beyond nearest-neighbor
distance $r_{\rm nn}$ into $H^{\rm dip}_{>r_{\rm nn}}$.

The first term is therefore that of an {\it effective} nearest-neighbor
pyrochlore Ising {\it antiferromagnet}~\cite{Anderson:1956},
  which is frustrated and possess
a Pauling entropy as long as $(J+5D)/3>0$, while
a four sublattice ``all-in/all-out'' long-range N\'eel order occurs
for $(J+5D)/3<0$. Monte Carlo simulations~\cite{Melko:2001,Melko:2004}
and mean-field theory~\cite{Gingras:2001} find that
$H^{\rm dip}_{>r_{\rm nn}}$ only slightly modifies the location
of the N\'eel to spin ice boundary from $(J+5D)/3=0$ to
$(J+4.53D)/3 \approx 0$.
In other words, the long range dipole-dipole interactions slightly {\it stabilize} the
N\'eel order to the detriment of the spin ice 
state~\cite{Hertog:2000,Melko:2001,Melko:2004}.
Using the $J/3$ and $5D/3$ values above for 
Dy$_2$Ti$_2$O$_7$ and Ho$_2$Ti$_2$O$_7$, we find that both materials
fulfill the criterion $(J+4.53D)/3>0$,
as well as the less accurate $(J+5D)/3>0$ criterion, in order 
to be characterized as dipolar spin ice systems.

Having established an ``order 0'' criterion,
$(J+5D)/3>0$, to determine whether a dipolar pyrochlore Ising system
exhibits spin ice behavior, we now have a sharp question in hand:
how is it that the long range $1/{r_{ij}}^3$ tail of
$H^{\rm dip}_{>r_{\rm nn}}$ does not (seemingly) lift the
extensive degeneracy created by the nearest-neighbor part?
To answer this question, we first turn to mean-field theory.

\subsection{Mean-field theory}
\label{Section:mft}

The general idea of the Ginzburg-Landau theory is  to determine when the
paramagnetic phase, where all the components of 
the local (on-site) magnetization, $m_i^{a,u}$ vanish,
become spontaneously unstable (i.e. critical) against
the development of nonzero $m_i^{a,u}$.
To proceed, we
follow here the approach laid out in
 Refs.~\cite{Gingras:2001,Reimers:MFT,Enjalran:2004}.
Consider a general bilinear spin Hamiltonian 
\begin{equation}
H = \frac{1}{2} \sum_{(i,j)} S_i^{a,u}
{\mathcal K}^{ab}_{uv}(i,j)  S_j^{b,v}
\end{equation}
where $S_i^{a,u}$ is the $u=x,y,z$ components of spin $\vec S_i^a$ 
on the $a^{\rm th}$ sublattice of the $i^{\rm th}$ primitive basis
and ${\mathcal K}^{ab}_{uv}(i,j)$ is a generalized spin-spin interaction. 
Making the mean-field ansatz
\begin{equation}
(S_i^{a,u} - m_i^{a,u} ) (S_i^{b,v} - m_j^{b,v} ) =0 \; ,
\end{equation}
where $m_i^{a,u}$ is the thermal average $\langle S_i^{a,u} \rangle $, 
allows to decouple $H$ above and write it as an effective one-particle problem
from which the free-energy, 
${\mathcal F}(\{m_i^{a,u}\})\equiv -\beta^{-1}\ln(Z)$,
where $Z={\rm Trace}[\exp(-\beta H)]$ is the partition function,
can be expanded as a Taylor series of the order parameters $m_i^{a,u}$.
Here we take $\beta=1/T$, where $T$ is the temperature and work in
units where the Boltzmann constant $k_{\rm B}=1$.
The leading term in ${\mathcal F}$ is quadratic in the $m_i^{a,u}$:
\begin{eqnarray}
{\mathcal F} & \approx & \frac{1}{2}\sum_{i,j} \sum_{a,b} \sum_{u,v}
m_{i}^{a,u} \left\{ nT\delta_{i,j}\delta^{a,b}\delta^{u,v}
- {\mathcal K}^{ab}_{uv}(i,j) \right\} m_{j}^{b,v} + {\mathcal F}_0(T)\nonumber \\
\label{eq-FEreal}
\end{eqnarray}
where ${\mathcal F}_0(T)$ is a $m_i^{a,u}$-independent temperature-dependent function
and $n$ is the number of spin components of $\vec S_i^a$.
We next introduce the Fourier transform representation of
$m_i^{a,u}$
\begin{eqnarray}
\label{eq-ftm}
m_{i}^{a,u} = \sum_{\vec q} m_{\vec q}^{a,u} {\rm e}^{-\imath {\vec q} \cdot {\vec R}_{i}^{a}} \ , \\
{\mathcal K}^{ab}_{uv}(i,j) = \frac{1}{N_{\rm cell}}
\sum_{\vec q} {\mathcal K}^{a b}_{uv}({\vec q}) {\rm e}^{\imath {\vec q} \cdot {\vec R}_{ij}^{ab}} \ ,
\label{eq-ftj}
\end{eqnarray}
where $N_{\rm cell}$ is the number 
of FCC Bravais lattice points, and ${\vec R}_i^a$ denotes
the position of a magnetic ion on sublattice $a$ of basis cell $i$. 
Equations (\ref{eq-ftm}) and (\ref{eq-ftj}) applied to ${\mathcal F}$ give
\begin{eqnarray}
\frac{({\mathcal F} - {{\mathcal F}}_0 )}{N_{\rm cell}} 
&=& \frac{1}{2}\sum_{\vec q} \sum_{a,b} \sum_{u,v}
m_{\vec q}^{a,u} \left( nT\delta^{a,b}\delta^{u,v}
- {\mathcal K}^{ab}_{uv}({\vec q}) \right) m_{-{\vec q}}^{b,v} \; .
\end{eqnarray}

To proceed, a transformation to normal modes is
necessary to diagonalize ${\mathcal K}^{ab}_{uv}({\vec q})$ via
\begin{equation}
\label{eq-normmodes}
m_{\vec q}^{a,u} = \sum_{\alpha=1}^4 \sum_{\mu=1}^3 U^{a,\alpha}_{u,\mu}({\vec q})
\phi_{\vec q}^{\alpha,\mu} \ ,
\end{equation}
where the Greek indices $(\alpha,\mu)$ label the normal modes,
$\phi_{\vec q}^{\alpha,\mu}$.
$U({\vec q})$ is the unitary matrix that diagonalizes
${\mathcal K}^{ab}_{uv}({\vec q})$ 
in the spin$\otimes$sublattice space with eigenvector $\lambda({\vec q})$,
\begin{equation}
\label{eq-unitaryT}
U^{\dagger}({\vec q}){\mathcal K}({\vec q})U({\vec q}) = \lambda({\vec q}) \ ,
\end{equation}
where, in component form, $U^{a,\alpha}_{u,\mu}({\vec q})$ 
represents the $(a,u)$ component of
the $(\alpha,\mu)$ eigenvector 
at ${\vec q}$ with eigenvalue $\lambda^{\alpha}_{\mu}({\vec q})$.
Therefore, the mean-field free energy, to quadratic 
order in the normal modes variables, is
\begin{eqnarray}
\label{eq-fenm}
{\mathcal F}(T) &\sim & \frac{1}{2}
\sum_{\vec q}\sum_{\alpha,\mu} \phi_{\vec q}^{\alpha,\mu} \left( nT -
\lambda^{\alpha}_{\mu}({\vec q}) \right) \phi_{-{\vec q}}^{\alpha,\mu} \; .
\end{eqnarray}

For the Ising case that interests us, we have $n=1$ and
 the indices representing the spin components $(u,v)$ 
and the label $\mu$ are dropped from Eq.~(\ref{eq-fenm}).
Our main goal is to diagonalize ${\mathcal K}^{ab}$ which,
by taking a tetrahedron primitive basis 
with local $\langle 111\rangle$ Ising spins, is
now simply a $4\times 4$ matrix,
 and determine the spectrum $\lambda^\alpha({\vec q})$.
Let us define $\lambda^{\rm max}({\vec q})$, the largest of the four 
$\lambda^\alpha({\vec q})$ at each ${\vec q}$.
Let us assume now that the
3-dimensional surface 
$\lambda^\alpha({\vec q})$ (as a function of $(q_x,q_y,q_z)$) 
reaches an absolute maximum value
over the whole Brillouins zone  for some 
$\vec q_{\rm ord}$, which we shall label 
$\lambda^{\rm max}({\vec q}_{\rm ord})$.
 ${\vec q}_{\rm ord}$ is called the {\it ordering wave vector} 
and $\lambda^{\rm max}(\vec q_{\rm ord})$ is the {\it critical temperature}, $T_c$,
because, when the temperature $T$ drops 
below $\lambda^{\rm max}(\vec q_{\rm ord})$, 
$\left(T -\lambda^{\rm max}({\vec q}_{\rm ord}) \right)$
in Eq.~(\ref{eq-fenm}) changes sign, causing the mode
$\phi_{{\vec q}_{\rm ord}}^{\rm max}$ to go soft (i.e. critical)
and to develop a nonzero thermal expectation value, at least
according to mean-field theory.
A key property of highly frustrated systems, 
such as the classical Heisenberg antiferromagnet on
pyrochlore, kagome and face-centered cubic lattices, 
is that an infinitely large number of modes go soft/critical 
simultaneously at $T_c$~\cite{Reimers:MFT}.

\begin{figure}[ht] 

\includegraphics[width=5.5cm]{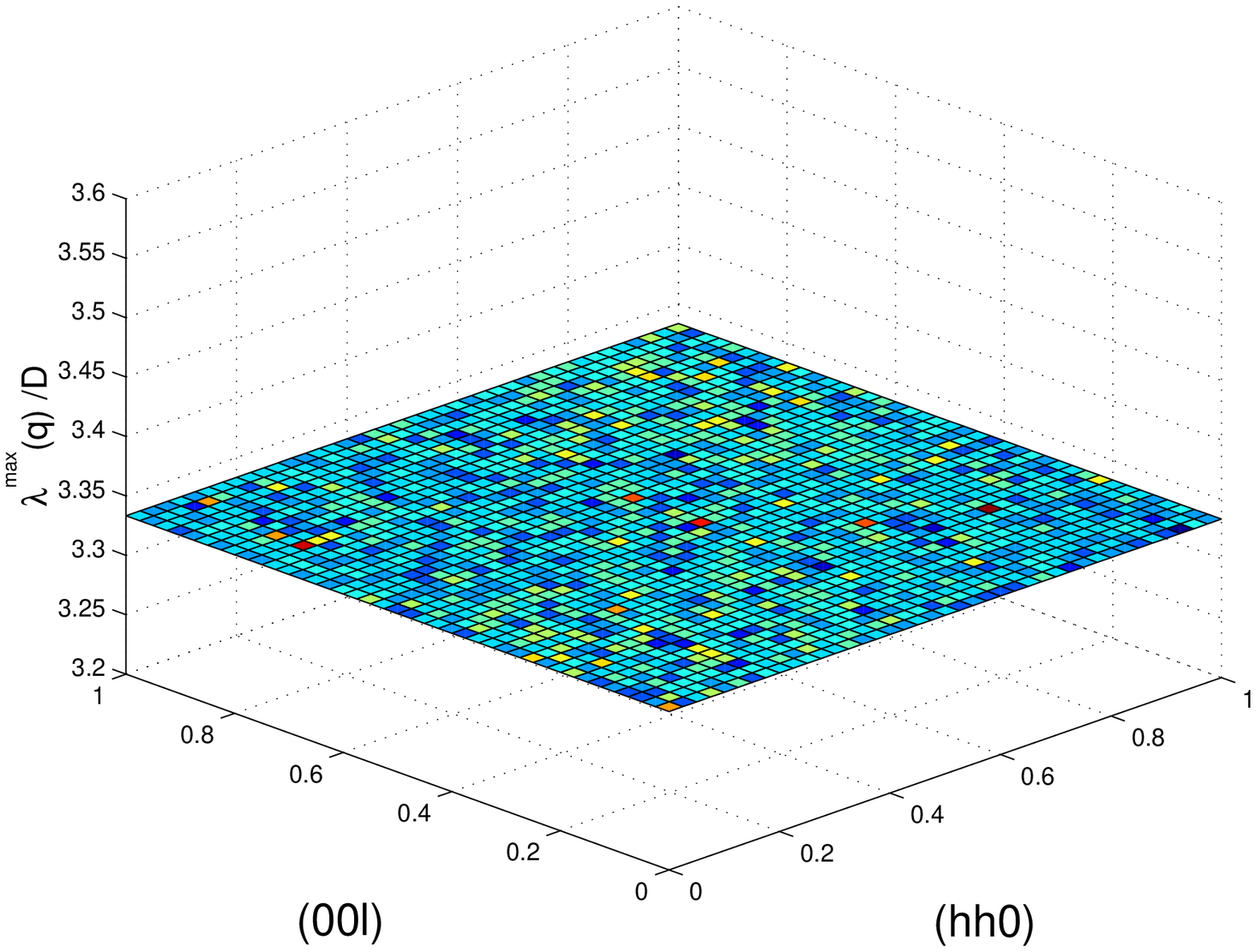}
\includegraphics[width=0.5cm]{sp}
\includegraphics[width=5.5cm]{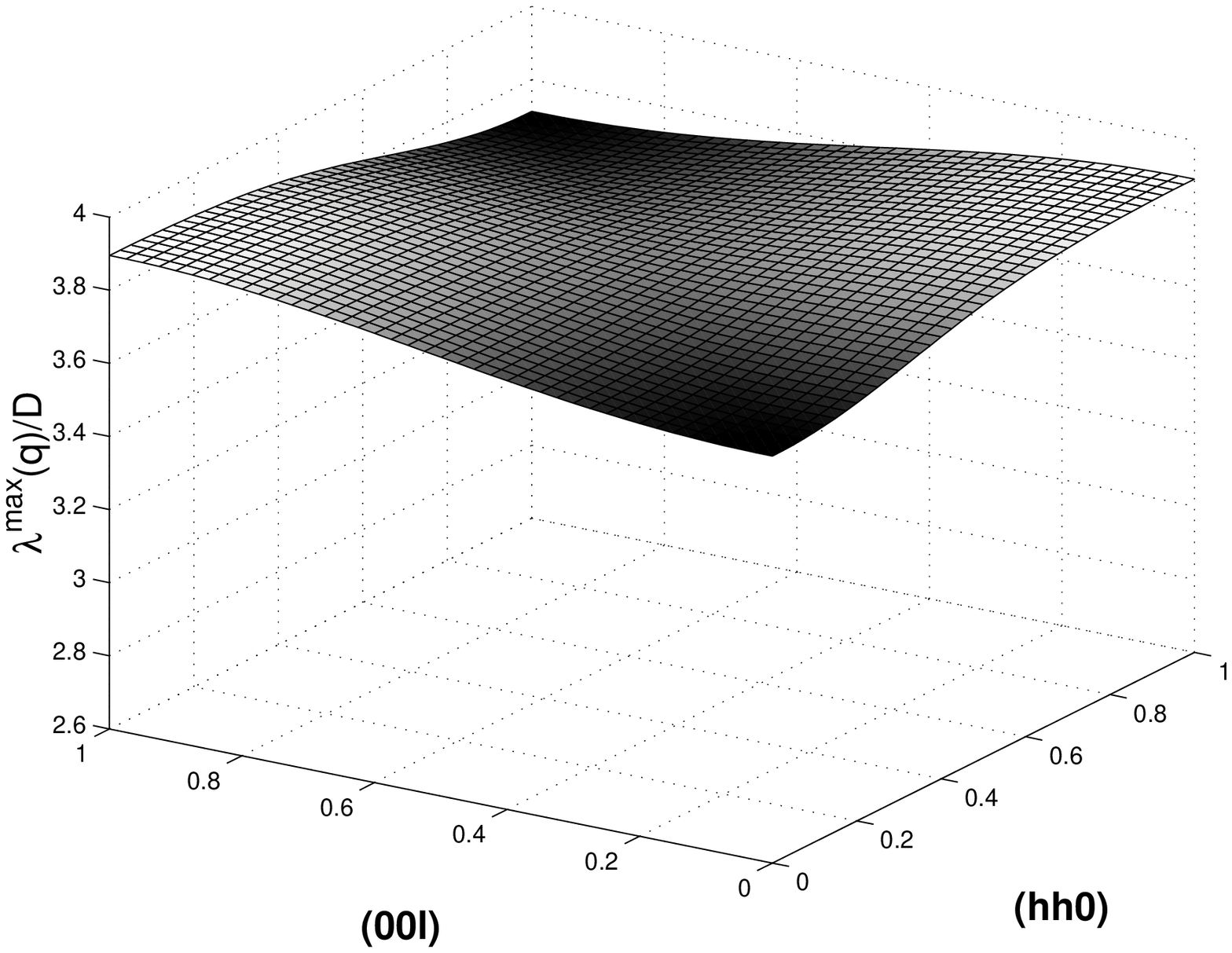}
\caption{Left:
 spectrum of $\lambda^{\rm max}({\vec q})/D$ for  an effective 
nearest-neighbor spin ice model 
where the dipole-dipole interactions are truncated at nearest-neighbor
distance.
The spectrum is flat and there is no selected ordering
wave vector ${\vec q}_{\rm ord}$.
Since there is one net uncompensated spin out of the three
spins neighboring a reference spin on a given tetrahedron,
the net ``mean-field''
acting at a site is twice the effective nearest-neighbor interaction
for the Ising variable, i.e. $5D/3$. Hence, the mean-field $T_c$ is
$10D/3\sim 3.33$ corresponding here to the ${\vec q}$-independent
$\lambda^{\rm max}({\vec q})$. 
Right:  $\lambda^{\rm max}({\vec q})/D$ for the full dipolar spin ice model 
with $J_{ij}=0$ in Eq.~(\ref{Hdsm}). Note the slightly dispersive
nature of  $\lambda^{\rm max}({\vec q})/D$
which displays a maximum at ${\vec q}_{\rm ord}=001$.}
\label{spectrum}
\end{figure}

For the $\langle 111\rangle$ nearest-neighbor 
Ising pyrochlore ferromagnet (first term on the right hand side of
Eq.~(\ref{SIcriterion})), or Anderson's pyrochlore
Ising antiferromagnet, one finds that there are two exactly degenerate 
branches of soft modes that have an identical eigenvalue $\lambda^{\rm max}(\vec q)$ 
for {\it each} wave vector ${\vec q}$ in the Brillouin zone (see Fig.~\ref{spectrum}).
This result is obtained after constructing the $4\times 4$ 
${\cal K}^{ab}({\vec q})$ matrix obtained from Eq.~(\ref{eq-ftj})
and finding its 4 eigenvalues. 
The mean-field theory of the nearest-neighbor model therefore predicts, 
via a reciprocal space description, that there
is no unique ordering wavevector ${\vec q}_{\rm ord}$ developping at
$T_c$. In fact, there are $2N_{\rm cell}$ modes 
(i.e. an extensive number of modes) that go soft simultaneously at
$T_c= 2(J/3+5D/3)$, where $N_{\rm cell}=N/4$ is the number of primitive 
basis cells, and $N$ is the number of magnetic moments.
This result agrees with the discussion that the nearest-neighbor spin ice model
has a degenerate ground state.

What about the role of the long range 
dipole-dipole interactions beyond nearest neighbors
contained
in $H^{\rm dip}_{> {\rm nn}}$ in Eq.~(\ref{SIcriterion})? 
Since we are still only considering an Ising 
model, the $4\times 4$ $K^{ab}({\vec q})$ matrix
is merely constructed by adding the
 Fourier transform of $H^{\rm dip}_{> {\rm nn}}$.
Because of the long-range $1/{r_{ij}}^3$ nature of the dipolar interactions,
it is somewhat technical to proceed with the calculation of
the Fourier transform in Eq.~(\ref{eq-ftj})
~\cite{Enjalran:2004,Isakov:2005}. 
The simplest, although not well controlled way to do so,
is to truncate $H^{\rm dip}_{> {\rm nn}}$ at some cut-off distance $R_c$ and determine
$\lambda^{\rm max}({\vec q}_{\rm ord})$ for that $R_c$~\cite{Gingras:2001}.
One can then monitor 
the evolution of the $\lambda^{\rm max}({\vec q})$ ``surface'' and the
corresponding ${\vec q}_{\rm ord}$ with progressively increasing $R_c$.
For ``small'' $R_c$, say $R_c$ 
less than the tenth nearest-neighbor distance, 
one finds that  $\lambda^{\rm max}({\vec q})$ acquires sizeable dispersion 
and  a ${\vec q}_{\rm ord}$ is selected. However, the value
of ${\vec q}_{\rm ord}$ depends on the chosen value of $R_c$, a
physically unacceptable result given that ultimately $R_c=\infty$ must
be considered~\cite{Gingras:2001}. 
As $R_c$ is pushed to larger and larger value, 
$\lambda^{\rm max}({\vec q})$ becomes progressively less dispersive with 
a ${\vec q}_{\rm ord}$ moving towards $0 0 1$. This result is
confirmed by calculations of ${\cal }K^{ab}(\vec q)$ 
directly in ${\vec q}$ space using the Ewald-summation 
method~\cite{Enjalran:2004v1}.
Figure~\ref{spectrum} shows the ${\vec q}$ dependence of 
$\lambda_{\rm max}({\vec q})$ in the $(hhl)$ reciprocal plane.

\newpage

One interpretation of the above mean-field theory calculation is
that there exists a sort of {\it self-screening} in place such
that the selection of an ordering wave vector 
${\vec q}_{\rm ord}$  from progressively increased 
cut-off radius $R_c$ cancels the ordering ``promoted''
by shells of spins of smaller radii. 
However, this is an incomplete
description of what really is the microscopic 
reason as to why the dipolar interactions give rise to a
fulfillment of the ice rules. 
In other words, why does dipolar spin ice obeys the ice rules~\cite{Isakov:2005}?
We return to this question in Section~\ref{Section:origin}. 
Let us for the moment just take the
results of the mean-field calculations at face value. 
These show that the ultimate selection of 
a  ${\vec q}_{\rm ord}$ is ``fragile''
in the dipolar spin ice model. 
By ``fragile'', we refer to the flatness of the spectrum of 
$\lambda^{\rm max}({\vec q})$ which shows that dipolar spin ice
has very little propensity to select an ordered state.
Indeed, a slight modification of the exchange interactions beyond
nearest-neighbors in Eq.~(\ref{Hdsm}) can dramatically alter the
spin-spin correlations in the spin ice regime~\cite{Yavorskii:2008}.
Yet, as mentioned above, a careful consideration
of $\lambda^{\rm max}({\vec q})$
does reveal that there is an absolute maximum of 
$\lambda^{\rm max}({\vec q})$ at $\vec q_{\rm ord}= 0 0 1$. In other
words, dipolar spin ice is characterized by an ordered phase 
with a {\it unique} commensurate propagation wave vector and that,
as the system is cooled down from the paramagnetic phase, 
it should undergo a phase transition to a long-range ordered ground state with no
extensive degeneracy. Since this is a classical system, quantum
fluctuations cannot inhibit the development of long-range order
as occurs in systems such as described in other chapters of this book.
Mean-field theory alone ``saves'' the third law of thermodynamics for
the dipolar spin ice model without having to invoke the effects
of quantum mechanics at low temperature.

What happens to 
this predicted long-range ordered phase in both the Monte Carlo simulations 
of the dipolar spin ice model and in real systems?
It is easier to first address this question in the case 
of the simulations, as we  do in the next section.

\subsection{Loop Monte Carlo simulations and phase diagram of dipolar spin ice}
\label{Section:loopMC}

In Monte Carlo simulations that employ simple conventional Metropolis single
spin flip dynamics~\cite{Hertog:2000,Bramwell:2001}, 
the rate of accepted spin flips dies off
exponentially fast with decreasing 
temperature once the system has
entered the spin ice regime where the ``two-in/two-out'' ice rules are
firmly obeyed~\cite{Melko:2004}. For example, in simulations of
Dy$_2$Ti$_2$O$_7$, one finds that it becomes for all practical purposes
essentially impossible to equilibrate the system below a ``freezing
temperature'' $T_f$ around 0.4 K~\cite{Melko:2004}. 
Interestingly, this temperature
more or less corresponds
 to the freezing temperature found in
AC susceptibility measurements on Dy$_2$Ti$_2$O$_7$~\cite{Fukazawa:2002}.
One can then re-interpret the results of mean-field theory in the light 
of the Monte Carlo results. The flatness of $\lambda^{\rm max}(\vec q)$
illustrates the competition between all the quasi-degenerate 
ice-rule obeying quasi-critical modes that exist in the dipolar spin ice model.
In real space, once the system is cooled into the ice regime,
that is below a temperature at which the specific heat peaks 
(see top panel in Fig.~\ref{Dy2Ti2O7-Cv}), it becomes
extremely difficult to flip spins that would violate the ice rules 
when the temperature is much lower than 
the energy barrier of approximately $2(J/3+5D/3)$.
Hence, while the low temperature spin dynamics within the spin ice state is
extremely sluggish, the system has not yet reached the critical temperature
to long-range order. The spin ice state, or regime, must therefore
be seen as smoothly connected with the high temperature paramagnetic phase.
In other words, the spin ice state is a paramagnetic state, a {\it 
collective
paramagnetic} state to be more precise and to
 borrow Villain's terminology~\cite{Villain:1979}.
That is, spin ice is a classical spin liquid, albeit with extraordinarily
slow spin dynamics. One is therefore led to ask a number of questions:
How can the
out-of-equilibrium freezing of spin ice be beaten, at least within
a computational attack on the dipolar spin ice model?
What is 
the long-range ordered phase, what is the nature of
the phase transition and at what temperature does it ultimately occur?

\begin{figure}[ht]
\begin{center}
\includegraphics[height=5cm,width=5cm,angle=90]{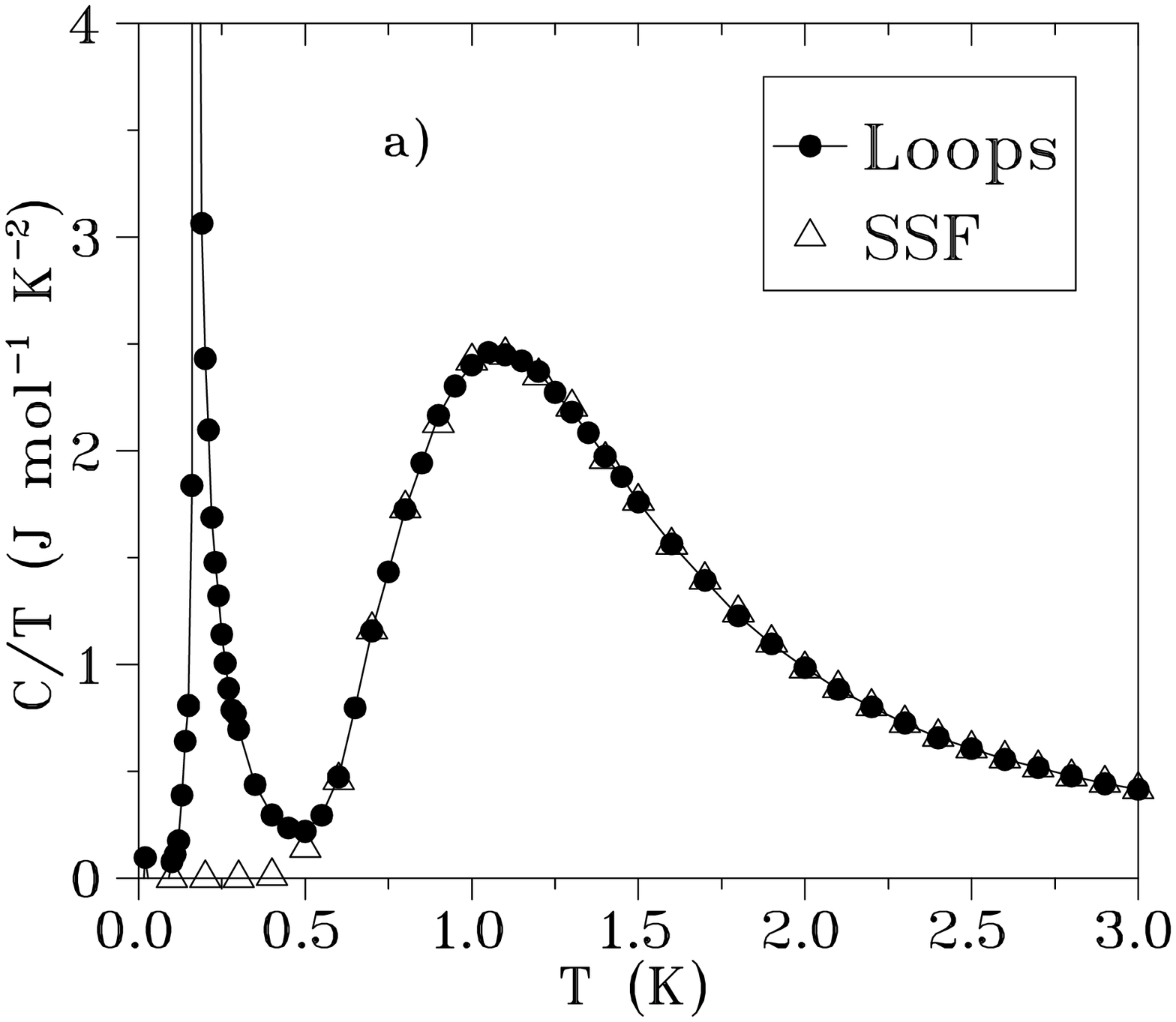}
\includegraphics[height=5cm,width=6cm,angle=0]{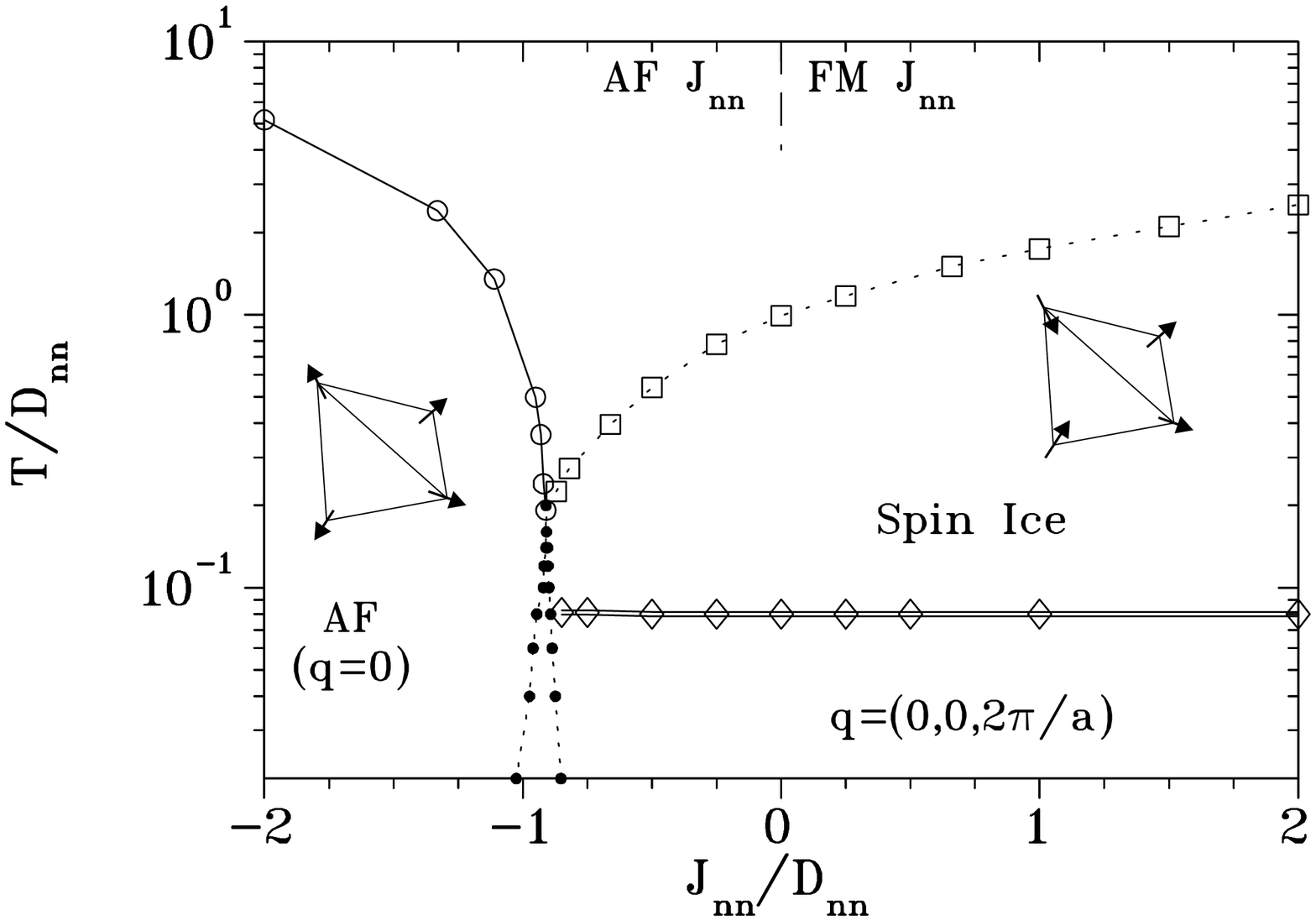}
\caption{Results from loop Monte Carlo simulations.
Left: the low temperature magnetic specific heat
for a system size $L=4$ and for $J/3=-1.24$~K and $5D/3=2.35$~K values
for Dy$_{\rm 2}$Ti$_{\rm 2}$O$_{\rm 7}$.
Closed circles are simulation data using
loop Monte Carlo simulations~\cite{Melko:2001,Melko:2004}.
Open triangles are data obtained using standard single spin
flip Metropolis algorithm~\cite{Hertog:2000}.
Right: Phase diagram of the dipolar spin ice model with
$J_{\rm nn}\equiv J/3$ and $D_{\rm  nn}=5D/3$.
The antiferromagnetic four sublattice N\'eel ground state is
with ``all-spins-in/all-spins-out'' configuration
for each tetrahedron.
The spin ice configuration, which includes the
${\bf q}=(0,0,2\pi/a)$ ground state, is a ``two spins in/two spins out''
configuration for each tetrahedron~\cite{Melko:2001,Melko:2004}.
The region encompassed between the quasi vertical dotted
lines displays hysteresis in the long-range ordered state selected
(${\bf q}=0$ vs. ${\bf q}=(0,0,2\pi/a)$) as
 $J_{\rm nn}/D_{\rm nn}$ is varied at fixed temperature $T$.}
\label{loopMC}
\end{center}
\end{figure}

In order to explore the low temperature ordering properties
of dipolar spin ice,
one needs a Monte Carlo algorithm with non-local updates that
effectively bypass the energy barriers that separate nearly degenerate
states and allows the simulation to explore the restricted ice-rules
phase space that prevents ordering in the model~\cite{Melko:2001,Melko:2004}.
One first identifies
the true zero energy modes that can take the near-neighbor spin ice model
from one ice state to another exactly energetically degenerate ice state.
An example of these zero modes, or loops, is shown in Fig.~\ref{pyrochlore}.
With interactions beyond nearest-neighbor, these ``loop moves'',
where spins are flipped on closed loops without violating the ice rules,
become {\it quasi-zero modes} that can take the
dipolar spin ice model from 
one ice-rules state to another without introducing ice-rule defects into the tetrahedra.
This allows all of the quasi-degenerate spin ice states to be sampled
ergodically, and facilitates the development of a long-range ordered state
at low temperatures.
The algoritm used to identify loops of ``potentially flipable spins'' is
described in Refs.~\cite{Melko:2001,Melko:2004,Barkema:1998,Barkema:book}.
Once a loop has been identified, the energy difference,
between the original spin configuration for that loop and the one with flipped spins,
which is due to the long-range part of the dipolar interactions,
is calculated. The flipping of the spins on the loop is then decided upon according
to a standard Metropolis test.
From the loop Monte Carlo simulations~\cite{Melko:2001,Melko:2004},
one finds, for $J/3=-1.24$~K and $5D/3=2.35$~K, 
a sharp first order transition
at $T_c\approx 180$~mK,
or $T_c/D \approx 0.128$ (see Fig.~\ref{loopMC}).
A detailed
analysis reveals several properties at the transition~\cite{Melko:2001,Melko:2004}:
\begin{enumerate}
\item The transition is an extremely strong first order transition.
\item The entropy removed on the high and low temperature ``wings'' of the transition
plus the latent heat associated with the transition equals Pauling's residual
entropy within 1\%$-$2\%.
\item The state below $T_c$ is a long-range ordered ice-rules obeying state 
and is identical to the one predicted by mean-field theory~\cite{Gingras:2001,Enjalran:2004v1}. 
\item $T_c$ is independent of $D/J$, with $T_c/D\approx 0.128$. 
\end{enumerate}

One can now return to the question asked earlier: 
``What happens to the numerically predicted phase transition
to long-range order in real systems?''
The numerical evidence for a first order transition to long-range order in
the dipolar spin ice model is compelling~\cite{Melko:2001,Melko:2004}. 
However, there has so far not been any reported experimental evidence for
a phase transition in either Dy$_2$Ti$_2$O$_7$~\cite{Fukazawa:2002}
or Ho$_2$Ti$_2$O$_7$~\cite{Harris:JMMM1998,BGH}  down to approximately 60 mK. 
A possible, if not likely, reason for this failure of
real spin ice materials to develop long range order at low temperature is that
the spins in real materials are frozen and cannot thermally equilibrate since 
real systems ``do not benefit'' from the non-local type of
spin dynamics such as the one employed in loop Monte Carlo simulations.
Another possibility~\cite{Yavorskii:2008} is that perturbative $J_{ij}$ exchange
interactions beyond nearest-neighbor in Eq.~(\ref{Hdsm}) frustrates the development of
the long range order discussed above, pushing $T_c$ below the lowest temperature considered
so far in experiments~\cite{Fukazawa:2002}.

Before we briefly discuss in the last section of this chapter some avenues
of research currently 
pursued in experimental studies of spin ices and related materials, 
we revisit the question of the microscopic origin of the ice rules 
in the dipolar spin ice model.

\subsection{Origin of ice rules in dipolar spin ice}
\label{Section:origin}

The Monte Carlo simulation results of Section~\ref{Section:competing-int}
show that the spin ice 
phenomenology is due to long-range dipolar interactions.
The mean-field theory calculations of Section~\ref{Section:mft} 
provide a first
clue as to the mechanism behind the formation of
the ice rules. However, neither method really explains why the
dipolar spin ice obeys the ice rules.
A succesful approach to answer this question was reported in Ref.~\cite{Isakov:2005}.

The first important ingredient~\cite{Isakov:2004} 
is to note that the  ``two-in/two-out'' ice rules on the pyrochlore lattice,
\begin{equation}
\sum_{\Delta} \sigma_{\Delta}^{z_1}
+\sigma_{\Delta}^{z_2}
+\sigma_{\Delta}^{z_3}
+\sigma_{\Delta}^{z_4}=0	\; ,
\end{equation}
where the sum is carried over up and down tetrahedra, is
equivalent to a divergence-ree ``spin field'' $\vec B$, $\nabla \cdot {\vec B}=0$.
These ``lattice flux'' are link variables that ``live'' on the diamond lattice dual
to the pyrochlore lattice and $\nabla\cdot$ is the lattice divergence.
By introducing a weight $\rho[{\vec B({\vec r})}]$
for nonzero flux (locally broken ice rules) of the form
\begin{equation}
\rho[{\vec B({\vec r})}] \propto \exp \left ( -\frac{K}{2}\int [\vec B({\vec r})]^2 d^3x \right )   \; ,
\end{equation}
one can solve for the correlations
\begin{eqnarray}
\langle B_u(0)\cdot B_v(\vec x)\rangle & \sim & \frac{3x_ux_v-r^2 \delta_{uv}}{r^5}
\label{gauge}
\end{eqnarray}
where $u,v$ are cartesian components.
Equation (\ref{gauge}) reveals that the local constraint, the ice rules,
lead to spin-spin
correlations that are dipolar-like at large distance.

The second important observation~\cite{Isakov:2005} 
it that one can construct a spin-ice-like
toy model whose eigenmodes  $\phi_{\vec q}^{\alpha}$
can be used to form a projector onto the ice-rule obeying ground states.
Because of the asymptotic dipolar-like correlations
that  ice-rules obeying ground state possesses, according to Eq.~(\ref{gauge}),
the {\it real space} matrix elements of this projector turns out
to be the same as the dipolar part of the interactions (the second term)
in $H_{\rm DSM}$ plus a small rapidly ($1/{r_{ij}}^5$) 
converging
correction term, $V_{ab}(r_{ij})$.
In other words, the dipolar interactions between the real magnetic moments
is, up to a small rapidly decaying short-range corrections, a {\it projector}
onto all possible ice-rule ground states. 
Hence, as a dipolar $\langle 111\rangle$ Ising system is cooled down below
a temperature of the order of $D$,
the spin configurations get energetically forced onto an ice-rule state, 
almost indiscriminently initially, when $V_{ab}(r_{ij}) < T < D$.
However, ultimately, the temperature reaches the scale of $V_{ab}(r_{ij})$
and a transition to a long-range ordered spin ice state, the one discussed
in Section~\ref{Section:loopMC}, occurs, 
provided ergodicity can be maintained~\cite{Melko:2001,Melko:2004}.

A possibly much simpler explanation has recently been proposed~\cite{Castelnovo:2007}.
``Exploding'' each point dipole into its constitutive magnetic (monopole) charges,
and imposing that each tetrahedron unit cell is neutral (magnetic charge wise),
automatically leads to the conclusion that all ice-rule obeying states are,
again up to a small rapidly decaying short-range correction, 
a ground state of the microscopic magnetostatic 
dipolar interactions of Eq.~(\ref{Hdsm})~\cite{Castelnovo:2007}.
Those magnetic charges interact as $1/r$ for large inter-charge separation $r$.
Magnetic field driven~\cite{Castelnovo:2007} and temperature driven~\cite{Jaubert:2009}
nucleation of those objects has been invoked to explain thermodynamic~\cite{Castelnovo:2007}
and dynamical~\cite{Jaubert:2009} properties of spin ice materials. 
We discuss below their possible experimental manifestation.


\section{Current Research Topics in Spin Ices and Related Materials}

Since the discovery of spin ice behavior in Ho$_2$Ti$_2$O$_7$ \cite{Harris:1997}
and Dy$_2$Ti$_2$O$_7$ \cite{Ramirez:Nature1999}, there has been a flory
of research activities aimed at exploring the interesting thermodynamic and magnetic
phenomena offered by spin ices and closely related systems.
We briefly review in this section some of the topics that are of current interest 
in the field of spin ice research.

\subsection{Magnetic field effects} 
Because of the
large magnetic moment of Dy$^{3+}$ and Ho$^{3+}$ in
spin ice materials, even moderately small magnetic field can induce dramatic effects.
Of particular interest is the ``pinning'' of the spins by a sufficently 
strong field on 
one~\cite{Castelnovo:2007,Moessner:2003,Tabata:2006,Fennell:2007}, 
two~\cite{Ruff:2005}, three~\cite{Ruff:2005,Higashinaka:2005}
or the four~\cite{Jaubert:2008} sites of a primitive tetrahedron cell,
 depending of the field orientation. 
This gives rise to interesting 
collective behaviors such as phase 
transitions to long range order~\cite{Ruff:2005,Higashinaka:2005},
topological Kasteleyn transitions~\cite{Moessner:2003,Fennell:2007,Jaubert:2008}, 
magnetization plateau states~\cite{Castelnovo:2007,Moessner:2003} and
a sort of magnetic monopole gas-liquid transition~\cite{Castelnovo:2007}.
We briefly discuss some of the salient features characterizing the behavior
of the system according to the direction of the magnetic field, $\vec B$.

\subsubsection{Field along $[\alpha,\beta,(\bar\alpha + \bar\beta)]$, $[11\bar 2]$}

The case of a very strong $\vec B$ along 
$\alpha \hat x + \beta \hat y -(\alpha+\beta)\hat z$ 
(e.g. $[11\bar 2]$ direction)
with $\alpha+\beta \ne 0$ 
(e.g. $[1 \bar 1 0]$) is perhaps the simplest one from the point of view
of field-induced collective behavior in dipolar spin ice systems~\cite{Ruff:2005}.
For $\vec B$ in that direction, the magnetic moments on sublattice \#1
with Ising easy-axis direction 
$\hat z_1 = \frac{1}{\sqrt 3}(\hat x + \hat y + \hat z)$ (see Section~\ref{Section:Hmic})
are perpendicular to $\vec B$
and are therefore decoupled from it. The moments on the three other sublattices 
are frozen (pinned) in a ``one-in/two-out'' configuration by a strong $\vec B$ in
that direction. In the limit $\vert \vec B\vert \rightarrow \infty$, the field-decoupled 
moments on sublattice \#1, which forms a regular FCC lattice,
are left to interact among themselves via dipolar and exchange interactions at and beyond
third nearest-neighbor distances. Numerical simulations~\cite{Ruff:2005} 
and experiments~\cite{Higashinaka:2005}  find that,
because of those interactions, the field-decoupled moments undergo a phase transition
to a ferromagnetic state in Dy$_2$Ti$_2$O$_7$ at approximately 0.3 K.

\subsubsection{Field along $[1 \bar 1 0]$}

The situation where $\vec B$ is along $[1 \bar 1 0]$ is a special case of the above
$[\alpha,\beta,(\bar\alpha + \bar\beta)]$ field direction.
For $\vec B$ along $[1 \bar 1 0]$, moments on sublattices \#1 and \#2, with
 $\hat z_1 = \frac{1}{\sqrt 3}(  \hat x + \hat y + \hat z)$ and
 $\hat z_2 = \frac{1}{\sqrt 3}(- \hat x - \hat y + \hat z)$, are decoupled from $\vec B$.
At the same time, the moments on sublattice \#3 and \#4, with
 $\hat z_3 = \frac{1}{\sqrt 3}(- \hat x + \hat y - \hat z)$ and
 $\hat z_4 = \frac{1}{\sqrt 3}(  \hat x - \hat y - \hat z)$, are pinned by a strong
$[1 \bar 1 0]$ field.
The pyrochlore lattice can be viewed as two sets 
of perpendicular chains, each set made of,
respectively, by the (\#1,\#2) sublattices, which form the so-called $\beta$ chains, 
and the (\#3,\#4) sublattices, which form the $\alpha$ chains~\cite{Hiroi:2003}
Here, one theoretically expects that the field-decoupled spins on the $\beta$ 
chains to undergo
a collective phase transition, again driven by dipolar and exchange interactions at and
beyond third nearest-neighbor distance~\cite{Ruff:2005,Yoshida:2004}.
 Results from specific heat~\cite{Hiroi:2003}
and neutron scattering measurements~\cite{Harris:1997,Fennell:2005,Clancy:2009} 
provide evidence that strong correlations develop 
among spins on the $\beta$ chains for even a moderate $[1 \bar 1 0]$ field 
$\vert \vec B\vert \sim 0.5$ T. However, in contrasts to theoretical predictions, 
no experiment has yet found true long range order among the $\beta$ chains.
Numerical~\cite{Melko:2004}
and experimental~\cite{Clancy:2009} evidence suggests that this failure to observe true
long range order may be due to small field misalignments 
away from perfect $[1 \bar 1 0]$.
Such an offset in the field direction 
may be sufficient to frustrate the three-dimensional
correlations among $\beta$ chains, 
causing the system to remain in a quasi-one-dimensional
short range state and failing to develop long range order.

\subsubsection{Field along $[111]$}

The pyrochlore lattice can also be viewed as an assembly of stacked kagome lattice planes
(i.e. planes of corner-shared triangles) separated by a triangular plane above and below
each kagome plane such that each triangle is decorated with spins alternatively 
above and below the kagome plane to give ``up'' and ``down'' tetrahedra.
At sufficiently low temperature, 
for a magnetic field $\vec B$ along $[111]$, the apical moments on each tetrahedron 
points along the field. That is, if $\vec B$ is ``up'', then spins point ``out''
on the ``up'' tetrahedra and ``in'' in the ``down'' tetrahedra.
For $\vert \vec B\vert$ less than a critical field value, $B_c$ of order of 1 T for
Dy$_2$Ti$_2$O$_7$ ~\cite{Sakakibara:2003}
and 1.7 T for Ho$_2$Ti$_2$O$_7$ ~\cite{Cornelius:2001,Fennell:2007}, 
the moments on the kagome plane
can still maintain ice-rule obeying configurations ``in-in-out'' or ``in-out-out''
on ``up'' and ``down'' tetrahedron, respectively. This leaves an intra-tetrahedron 
degree of freedom associated to the kagome planes (the apical moments are
polarized by the field) analogous to that in Pauling's model which, therefore, 
possess a residual zero point entropy. 
As long as $B<B_c$, this state with residual entropy,
referred to as {\it kagome ice}, is characterized by a
field-independent $[111]$ magnetization 
plateau~\cite{Cornelius:2001,Moessner:2003,Sakakibara:2003}
The magnetic entropy has been predicted~\cite{Moessner:2003}
and experimentally~\cite{Aoki:2004} found to be
non-monotonic as a function of $\vert \vec B\vert$
and to exhibit a spike at $B_c$. 
Very recent work interprets the transition from the plateau regime 
to the high field saturated magnetization regime
above $B_c$ as a condensation of magnetic monopole like defects~\cite{Castelnovo:2007}.
For $B<B_c$, the available degrees of freedom that live on the kagome planes can
be mapped to hard core dimers on a honeycomb lattice.  
Using this description, the kagome ice state is found to be critical.
Theory predicts that a tilting of the magnetic field
away from perfect $[111]$ alignment allows to 
control the entropy of this critical state, 
which is ultimately terminated for a tilting angle above a critical value 
by a topological phase transition called {\it Kasteleyn transition}~\cite{Moessner:2003}.
Neutron scattering intensity data on Ho$_2$Ti$_2$O$_7$ subject to
a magnetic field tilted away from $[111]$ 
have been compellingly interpreted in terms of such a Kasteleyn transition 
out of a critical kagome ice state~\cite{Fennell:2007}.

\subsubsection{Field along $[100]$}

A field $\vec B$ along $[100]$ has an equal projection along each of the $\hat z_i$ Ising 
easy axis. A field $\vert \vec B\vert$ larger than 0.04 T overwhelms the correlations cause
by the long range magnetostatic dipolar interactions and gives rise to a polarized
state along $[100]$ that satisfies the ice rules~\cite{Melko:2004}.
On the basis of Monte Carlo simulations, it was first believed
that for a field less than a critical value, the system undergoes a first order
liquid-gas transition between a state of low $[100]$ magnetization to a state with
higher $[100]$ magnetization upon cooling~\cite{Harris:1998}.
However, a more recent study shows that this problem is more subtle~\cite{Jaubert:2008}.
Indeed, at low temperature,
the approach to the saturated magnetization is another example of
a Kasteleyn transition which is topological in nature since
the magnetization can only be changed via correlated
spin excitations that span the whole system.
Comparison of simulation data with the
magnetic field dependence of
the $[100]$ magnetization for Dy$_2$Ti$_2$O$_7$ does provide
good evidence that the physics of the 
Kasteleyn transition is indeed at play~\cite{Jaubert:2008}.

\subsection{Dynamical properties and role of disorder}

Experiments have been carried out to investigate and unravel the nature of
the persistent spin dynamics observed down to the lowest 
temperature~\cite{Ehlers:2006,Sutter:2007,Lago:2007}.
The substitution of the magnetic Ho$^{3+}$ and Dy$^{3+}$ ions by non-magnetic
ions affects the spin dynamics and lead to a partial and {\it non-monotonous} 
lifting of the ground state degeneracy~\cite{Ke:2007}.
Yet, the interesting problem of the ultimate transition from
spin ice to dipolar spin glass as
the magnetic rare earth ion is substituted by Y$^{3+}$ has not yet been explored.
The origin of the spin dynamics in spin ices is not fully understood~\cite{Lago:2007}.
Recent work on Ho$_2$Ti$_2$O$_7$ proposes that the excitation of nuclear states
perturb the electronic Ising spin states and give
rise  to persistent spin dynamics~\cite{Ehlers:2009}.
A rather interesting aspect of the spin dynamics in spin ice materials,
such as Dy$_2$Ti$_2$O$_7$, is the 
faily temperature, $T$, independent relaxation rate $\tau(T)$ 
between 4 K and 13 K~\cite{Snyder:2003,Snyder:2004}.
This temperature independent regime had previously been interpreted in 
terms of quantum mechanical tunneling of the spins between their 
``in'' and ``out''  directions. However, recent work suggests that the 
magnetic relaxation can be entirely interpreted in terms of
the diffusion of thermally-nucleated topological defects,
again akin to magnetic monopoles, on trajectories constrained 
to lie on a network of ``Dirac strings''~\cite{Jaubert:2009}.

\subsection{Beyond the dipolar spin ice model}

The large number of high quality experiments on spin ices
are now permitting to refine the spin Hamiltonian and extract
the exchange interactions beyond nearest-neighbors, allowing an excellent
quantitative description of many bulk measurements and neutron scattering
experiments~\cite{Yavorskii:2008,Tabata:2006}.
In particular, exchange interactions beyond nearest-neighbors are argued to
induce weak spin-spin correlations that are responsible for the neutron
scattering intensity spanning the Brillouin zone boundaries~\cite{Yavorskii:2008}.

\subsection{Metallic spin ice}
The metallic Pr$_2$Ir$_2$O$_7$ pyrochlore material 
exhibits interesting Kondo-like effects with 
a logarithmic increase of the resistitivy and magnetic susceptibility
at low temperatures~\cite{Nakatsuji:2006,Machida:2007}.
One expects Pr$^{3+}$ in Pr$_2$Ir$_2$O$_7$ 
to be well described by an Ising spin, for which Kondo physics is not straightforwardly
explainable. Furthermore, the predominant Pr$^{3+}$$-$Pr$^{3+}$ interactions in
metallic Pr$_2$Ir$_2$O$_7$ would naively be expected to be RKKY-like. 
In contrast to experimental findings, Monte Carlo simulations on
RKKY-coupled $\langle 111\rangle$ spins find a transition to long range order
at a temperature of the order of the Curie-Weiss temperature~\cite{Kawamura:2008}.
Hence it is not really clear what is, if any, the role of magnetic 
frustration in determining the exotic transport and thermodynamic
 properties of Pr$_2$Ir$_2$O$_7$.

\subsection{Artificial spin ice}
Lithographically fabricated single-domain ferromagnetic islands can be arranged such that
the magnetic dipolar interactions create a two-dimensional analogue to spin ice.
The magnetic moments can be directly imaged, allowing to study the local 
accomodation of frustration and emergence of ice-rules~\cite{Wang:2006,Nisoli:2007}.
	Also, it has been proposed that a colloidal version 
of artificial ice can be realized using optical trap lattices~\cite{Libal:2006}.
It will be interesting to see more such artificially systems becoming available.
The out-of-equilibrium properties of the lithographically made systems subject
to nonzero magnetic field promise to be quite interesting.

\subsection{Stuffed spin ice}
One can chemically alter Ho$_2$Ti$_2$O$_7$ 
spin ice by ``stuffing'' extra  Ho$^{3+}$  
magnetic moments into 
the normally non-magnetic Ti sites~\cite{Lau1,Lau2}. 
 The resulting series, Ho$_2$(Ti$_{2-x}$Ho$_x$)O$_{7-x/2}$ 
displays an increased connectivity compared 
to the standard pyrochlore, making interesting
 the question as to how the Pauling entropy
of ``normal'' spin ice evolves with stuffing~\cite{Lau1,Lau3}.
At this time, the question of homogeneity of stuffed
spin ice materials remains to be ascertained and more work remains
to be done on these systems.

\subsection{Quantum mechanics, dynamics and order in spin ices}

The Tb$_2$Ti$_2$O$_7$~\cite{Gingras:2000,Mirebeau:2007}
 and Tb$_2$Sn$_2$O$_7$~\cite{Mirebeau:2007} 
 should both possess an Ising
ground state doublet, similarly to their Ho and Dy counterpart~\cite{Rosenkranz:2000}.
However, Tb$_2$Ti$_2$O$_7$ remains in a collective paramagnetic (spin liquid) state
down to 50 mK despite an antiferromagnetic Curie-Weiss temperature
 of -14 K~\cite{Gardner:1999,Gingras:2000} and
is seemingly not a spin ice state~\cite{Molavian:2007}.
On the other hand, neutron scattering~\cite{Mirebeau:2005} 
finds a long-range ordered spin ice configuration
in Tb$_2$Sn$_2$O$_7$, although muon spin relaxation studies find significant spin dynamics
in this compound~\cite{Dalmas:2006,Bert:2006,Giblin:2008}.
What is the role of the deviation from a
strictly Ising like description of the magnetic moments 
in these two materials and the
consequential quantum fluctuations is an interesting question that is not yet
resolved~\cite{Molavian:2007}.
In that context, it is interesting to note that the 
Pr$_2$Sn$_2$O$_7$ pyrochlore material has been proposed to 
display a ``dynamical spin ice state''~\cite{Zhou:2008}.
Little is known about the microscopic physics at play in this system.

\section{Conclusion}

Spin ice materials were discovered just over ten years ago. 
Since then, their experimental and theoretical investigations 
have raised many interesting questions regarding several aspects 
of the physics of frustrated magnetic systems.
At the same time, the study of spin ices 
have led to an enhanced global understanding of many 
fundamental issues pertaining to frustration in condensed matter
systems. 
Yet, there are still several unanswered questions, 
in particular in the context of low-temperature spin dynamics, random disorder 
and properties of metallic Ising pyrochlore systems.  
These will likely continue to attract attention for many years to come.

\section{Acknowledgements}
I thank Steve Bramwell, Benjamin Canals, Adrian Del Maestro,
Byron den Hertog, Sarah Dunsiger,
Matt Enjalran, Tom Fennell, Jason Gardner,
Bruce Gaulin, John Greedan,
 Ying-Jer Kao, Roger Melko, Hamid Molavian, Jacob Ruff
and Taras Yavors'kii for their 
collaboration on projects related to spin ice.
I acknowledge many useful 
and stimulating conversations over the past few years with 
Tom Devereaux,
Rob Hill, 
Rob Kiefl,
Jan Kycia,
Chris Henley, 
Peter Holdsworth, 
Graeme Luke, 
Paul McClarty,
Roderich Moessner, 
Jeffrey Quilliam, 
Peter Schiffer, 
Andreas Sorge,
Ka-Ming Tam,
Oleg Tchernyshyov
and
Jordan Thompson.
Funding from the NSERC of Canada, the Canada Research Chair Program,
the Research Corporation, the Canadian Institute for Advanced Research, 
the Ontario Innovation Trust,
the Ontario Challenge Fund and Materials and Manufacturing Ontario 
is gratefully acknowledged.
The hospitality of the Kavli Institute of Theoretical Physics 
(KITP) at the University of California of Santa Barbara, 
where part of this review was written, is acknowledged.
Work at the KITP was 
supported in part by the National Science Foundation under
Grant No. NSF PHY05-51164.

\printindex
\end{document}